\pdfoutput=1
\documentclass[12pt, a4paper, oneside]{article}
\usepackage{latexsym}
\usepackage{amssymb}
\usepackage{mathrsfs}
\usepackage{graphicx}
\usepackage[latin1]{inputenc}
\usepackage{natbib}
\usepackage{multirow}
\usepackage{url}
\usepackage{dcolumn}
\usepackage{amsthm}
\usepackage{amsmath}
\usepackage{array}
\usepackage{amsfonts}
\usepackage{bm}
\usepackage[german]{varioref}
\usepackage{psfrag}
\usepackage{caption}
\usepackage{rotating}
\usepackage{oldgerm}
\usepackage{enumerate}
\usepackage{parskip}
\usepackage{setspace}
\usepackage{hyperref}
\usepackage{here}	
\usepackage{pdflscape}
\usepackage{listings}
\usepackage{subcaption}
\usepackage[english]{babel}
\usepackage[normalem]{ulem}
\usepackage[lined,boxed,linesnumbered]{algorithm2e}
\usepackage{color}
\usepackage{footmisc}
\usepackage{enumitem}
\usepackage{dsfont} 
\usepackage{tabularx}
\usepackage{longtable}

\newcolumntype{L}[1]{>{\raggedright\arraybackslash}p{#1}} 
\newcolumntype{C}[1]{>{\centering\arraybackslash}p{#1}} 
\newcolumntype{R}[1]{>{\raggedleft\arraybackslash}p{#1}} 

\long\def\symbolfootnote[#1]#2{\begingroup%
\def\thefootnote{\fnsymbol{footnote}}\footnote[#1]{#2}\endgroup}

\numberwithin{equation}{section}

\newtheorem{Theorem}{Theorem}[section]

\newtheorem{Definition}[Theorem]{Definition}

\newtheorem{example}[Theorem]{Example}

\allowdisplaybreaks[1]
\newcommand{\BBB}{\mathcal{B}}
\newcommand{\CCC}{\mathcal{C}}

\newcommand{\HHH}{\mathcal{H}}

\newcommand{\LLL}{\mathcal{L}}
\newcommand{\NNN}{\mathcal{N}}

\newcommand{\RRR}{\mathcal{R}}
\newcommand{\SSS}{\mathcal{S}}

\newcommand{\UUU}{\mathcal{U}}
\newcommand{\VVV}{\mathcal{V}}
\newcommand{\WWW}{\mathcal{W}}
\newcommand{\coloneqq}{:=}

\DeclareMathOperator{\Variance}{Var}           

\DeclareMathOperator{\diag}{diag}
\begin{document}\sloppy

\title{\textbf{Selection of Sparse Vine Copulas in High Dimensions with the Lasso}}
\author{Dominik M\"uller\thanks{Corresponding author} \thanks{Department of Mathematics, Technische Universit\"at M\"unchen,
		Boltzmannstraße 3, 85748 Garching, Germany. E-Mail: \href{mailto:dominik.mueller@ma.tum.de}{dominik.mueller@ma.tum.de}, \href{mailto:cczado@ma.tum.de}{cczado@ma.tum.de}.}
	\quad
	Claudia Czado\footnotemark[2]}

\maketitle


\begin{abstract}
We propose a novel structure selection method for high dimensional ($d > 100$) sparse vine copulas. Current sequential greedy approaches for structure selection require calculating spanning trees in hundreds of dimensions and fitting the pair copulas and their parameters iteratively throughout the structure selection process. Our method uses a connection between the vine and structural equation models (SEMs). The later can be estimated very fast using the Lasso, also in very high dimensions, to obtain sparse models. Thus, we obtain a structure estimate independently of the chosen pair copulas and parameters. Additionally, we define the novel concept of regularization paths for R-vine matrices. It relates sparsity of the vine copula model in terms of independence copulas to a penalization coefficient in the structural equation models. We illustrate our approach and provide many numerical examples. These include simulations and data applications in high dimensions, showing the superiority of our approach to other existing methods.
\end{abstract}

\noindent%
{\it Keywords:}  Dependence Modeling, Vine Copula, Lasso, Sparsity


\section{Introduction}\label{sec:introduction}
Modeling dependence in high dimensional systems has become an increasingly important topic nowadays. This is mainly because data is more available but also computation capacities increase permanently. Hence, modeling joint distributions in arbitrary dimensions is key to understand and predict multivariate phenomena. Since analytically tractable multivariate distributions for arbitrary dimensions are hard to find and impose the same distributions on both marginals and dependency, copula models have become popular in recent decades. Based on the theorem of \cite{Sklar1959}, they enable modeling marginal distributions and dependency behaviour separately. This however only translates the problem of complex $d$-dimensional distributions to $d$-dimensional copulas. To overcome this, the pair copula construction (PCC) of \cite{Aasetal2009} allows for more flexible $d$ dimensional models. They consist of the marginal distributions and (conditional) bivariate copulas as building blocks, all of which can be chosen independently from each other. The resulting models, called \textit{regular vines} or \textit{R-vines} \cite{kuro:joe:2010} are specified by a sequence of $d-1$ linked trees, the R-vine structure, where the edges of the trees identify bivariate copulas. This method has been very popular in the last years in the financial context, see \cite{econometrics4040043} for an overview of applications. When it comes to determining a suitable R-vine structure, most often the algorithm of Dissmann \cite{dissmann-etal} is used. This locally greedy approach works well in lower dimensional setups. However, for high dimensional data it can not be ensured that its solutions are close to the optimum solution of this high dimensional combinatorial optimization problem. Our goal is to contribute another entirely different method for looking at the R-vine structure, scaling to hundreds of dimensions. This is necessary since e.\,g.\ the current Bayesian approaches \cite{GruberCzado20152}, \cite{GruberCzado2015} are computationally highly intensive and can not be used for more than $d \approx 20$ dimensions. Also, the Pair-Copula Bayesian Networks of \cite{BauerCzado2016} are not applicable in dimensions which exceed $d \approx 20$. Even though they exploit conditional independences given by a graphical model, they may ultimately involve high dimensional numerical integration. This is clearly a drawback to the pair-copula construction, which does not require integration at all. The work of \cite{MuellerCzado2016} proved to be several times faster than Dissmann's algorithm in moderately high dimensions, e.\,g.\ $d \approx 100$ by exploiting sparsity induced by DAGs modelled with a multivariate Gaussian distribution. As our approach, they also split the estimation of the R-vine structure from the pair copula estimates. Their approach to fit several DAGs with different degrees of sparsity has the drawback that each DAG generates a different R-vine structure. Thus, the fitting procedure has to be redone for each degree of sparsity, as with Dissmann's algorithm. Additionally, it still relies on maximum spanning trees.\\
The goal of this paper is to develop a novel approach exploiting in particular sparse structures. For this, we utilize the \textit{Lasso} \cite{Tibshirani94regressionshrinkage} which heavily influenced statistics in recent years by performing parameter estimation and model selection simultaneously. Introduced in the regression domain, it found widespread applications in other areas, such as the \textit{graphical Lasso} for graphical models, see \cite{glasso} and others. A very favourable property of the Lasso is the \textit{regularization path}, linking the Lasso-solutions to a tuning parameter $\lambda$, describing the degree of penalization for the respective solution.\\
Our approach relates vine copula models to structural equation models (SEMs) as introduced by \cite{BrechmannJoe2014}. This allows us to tap into the Lasso world by introducing a penalized regression on the structural equations which reflects the necessary properties for vine copula models, the so called \textit{proximity condition}. We show that Lasso-solutions to these structural equations, i.\,e.\ the regularization path, can be related to specific entries in the R-vine structure. By virtue of these concepts, we are able to introduce a regularization path concept for the R-vine itself. Thus, we obtain a high dimensional vine copula with a sparsity pattern reflecting the chosen degree of penalization.\\
The structure of the paper is as follows. First, we briefly introduce dependence modeling with R-vines in Section \ref{sec:rvines}. We  sketch the connection to structural equation models, which enables us to use the Lasso in Section \ref{sec:sem} and the Lasso will be reviewed in Section \ref{sec:lasso}. In Section \ref{sec:structureselectionLasso}, we introduce our novel approach by first considering the first R-vine tree and all subsequently estimated higher trees. We will define the \textit{R-vine regularization path} and discuss the choice of the tuning parameter $\lambda$, which controls the strength of penalization. In Section \ref{sec:numericalexamples}, we compare our approach to Dissmann's algorithm in a simulation study to show that our method deals better with sparse situations, especially present in high dimensional setups. After that, an example and outlook in $d > 200$ dimensions follows. We conclude the paper with a discussion in Section \ref{sec:discussion}.

\section{Dependence Modeling with R-vines}\label{sec:rvines}
We use the following conventions. Upper case letters $X$ denote random variables, and lower case letters $x$ their realizations. Bold lower case letters $\bm{v}$ denote vectors and bold upper case letters $\bm{M}$ denote matrices. Referring to sub-vectors, we denote by $v_i$ the $i$-th entry of the vector $\bm{v}$ and $\bm{v_{1:d}}$ the first $d$ entries of the vector $\bm{v}$. When considering matrices, we denote $m_{i,j}$ the $j$-th entry in the $i$-th row of the matrix $\bm{M}$. For rows or columns of a $d \times d$ matrix $\bm{M}$, we write $\bm{M}_{,j} = \left(m_{1,j},\dots,m_{d,j}\right)$ for the $j$-th column and\\ $\bm{M}_{i,} = \left(m_{i,1},\dots,m_{i,d}\right)$ for the $i$-th row of $\bm{M}$, respectively. Additionally, we have the following three data scales when working with copulas.
\begin{enumerate}\label{en:scales}
	\item \textit{x-scale}: the original scale of $X_i$, i.i.d., with density $f_i(x_i),\ i=1,\dots,d$,
	\item \textit{u-scale} or \textit{copula-scale}: $U_i = F_i\left(X_i\right)$, $F_i$ the cdf of $X_i$ and $U_i \sim \UUU\left[0,1\right]$, $i=1,\dots,d$,
	\item \textit{z-scale}: $Z_i = \Phi^{-1}\left(U_i\right)$, $\Phi$ the cdf of $\NNN\left(0,1\right)$ thus $Z_i \sim \NNN\left(0,1\right)$, $i=1,\dots,d$.
\end{enumerate}
We assume a random vector $\left(X_1,\ldots,X_d\right)$ with joint density function $f$ and joint distribution function $F$. By \cite{Sklar1959}, we can separate the univariate marginal distribution functions $F_1,\ldots,F_d$ from the dependency structure such that $F\left(x_1,\ldots,x_d\right) = \CCC\left(F_1\left(x_1\right),\ldots,F_d\left(x_d\right)\right)$,
where $\CCC$ is an appropriate $d$-dimensional copula. If $F_i$ are continuous, $\CCC$ is unique. The corresponding joint density function $f$ is given by
\begin{equation}\label{eq:copuladensity}
	f\left(x_1,\ldots,x_d\right) = \prod_{i=1}^d~f_i\left(x_i\right) \times c\left(F_1\left(x_1\right),\ldots,F_d\left(x_d\right)\right),
\end{equation}
where $c$ is a $d$-dimensional copula density. This expression incorporates a, possibly complex, $d$-dimensional copula density. As shown by \cite{Aasetal2009}, $d$-dimensional copula densities may be decomposed into $d\left(d-1\right)/2$ bivariate (conditional) copula densities. Its elements, the \textit{pair copulas} can be chosen completely independent from each other and display e.\,g.\, positive or negative tail dependence or asymmetric dependence. For a \textit{pair-copula-construction} (PCC) in $d$ dimensions, there exist many possible decompositions. These may be organized to represent a valid joint density by \textit{regular vines} (R-vines), see \cite{BedfordCooke2001, BedfordCooke2002}. A \textit{vine tree sequence} stores which bivariate (conditional) copula densities occur in the factorization of a $d$-dimensional copula density. Such a sequence in $d$ dimensions is given by $\VVV = \left(T_1,\ldots,T_{d-1}\right)$ such that
\begin{enumerate}[label={(\roman*})]
	\item $T_1$ is a tree with nodes $V_1=\left\{1,\ldots,d\right\}$ and edges $E_1$,
	\item for $i \ge 2$, $T_i$ is a tree with nodes $V_i = E_{i-1}$ and edges $E_i$,
	\item if two nodes in $T_{i+1}$ are joined by an edge, the corresponding edges in $T_i$ must share a common node (proximity condition (pc)).
	\label{eq:proximitycondition}
\end{enumerate}
To formalize this, define the \textit{complete union} $A_e$ of an edge $e$ by\\ $A_e \coloneqq \left\{j \in V_1|\exists \ e_1 \in E_1,\ldots,e_{i-1}\in E_{i-1}: j \in e_1 \in \ldots \in e_{i-1} \in e\right\}$ where the \textit{conditioning set} of an edge $e=\left\{a,b\right\}$ is defined as $D_e \coloneqq A_a \cap A_b$ and
$C_e \coloneqq C_{e,a} \cup C_{e,b} \mbox{ with } C_{e,a} \coloneqq A_a \setminus D_e \mbox{ and } C_{e,b} \coloneqq A_b \setminus D_e$ is the \textit{conditioned set}. Since $C_{e,a}$ and $C_{e,b}$ are singletons, $C_e$ is a doubleton for each $e,a,b$, see \cite[p.\ 96]{KurowickaCooke2006}. For edges $e \in E_i,\ 1 \le i \le d-1$, we define the set of bivariate copula densities corresponding to $j\left(e\right),\ell\left(e\right)|D\left(e\right)$ by $\mathcal{B}\left(\VVV\right) = \left\{c_{j\left(e\right),\ell\left(e\right);D\left(e\right)}|e \in E_i, 1 \le i \le d-1\right\}$ with the conditioned set $j\left(e\right),\ell\left(e\right)$ and the conditioning set $D\left(e\right)$. Denote sub vectors of $\mathbf{x}=\left(x_1,\ldots,x_d\right)^T$ by $\mathbf{x}_{D\left(e\right)} \coloneqq \left(\mathbf{x}_j\right)_{j \in D\left(e\right)}$. With the PCC, Equation \eqref{eq:copuladensity} yields
\begin{equation}\label{eq:vinedensity}
f\left(x_1,\ldots,x_d\right) =	\prod_{i=1}^{d}~f_i\left(x_i\right)
	\times
	\prod_{i=1}^{d-1}~\prod_{e\in E_i}~c_{j\left(e\right),\ell\left(e\right);D\left(e\right)}\bigg(F\left(x_{j\left(e\right)}|\bm{x}_{D\left(e\right)}\right),F\left(x_{\ell\left(e\right)}|\bm{x}_{D\left(e\right)}\right)\bigg).
\end{equation}
When we speak of bivariate \textit{conditional} copulas, we take into account the \textit{simplifying assumption}, which is imposing that the two-dimensional conditional copula density $c_{13;2}\left(F_{1|2}\left(x_1|x_2\right),F_{3|2}\left(x_3|x_2\right);x_2\right)$
is independent of the conditioning value $X_2 = x_2$ \citep{stoeber-vines}. The parameters of the bivariate copula densities  $\mathcal{B}\left(\VVV\right)$ are given by $\theta\left(\mathcal{B}\left(V\right)\right)$. This determines the R-vine copula $\left(\VVV,\mathcal{B}\left(\VVV\right),\theta\left(\mathcal{B}\left(\VVV\right)\right)\right)$. 
A representation of such a R-vine copula is most easily given by lower triangular $d\times d$ matrices, see \cite{dissmann-etal}. Such an R-vine matrix $\bm{M} = \left(m_{i,j}\right)_{i=1,\dots,d; j=1,\dots,d}$ has to satisfy three properties.
\begin{enumerate}[label={(\roman*})]
	\item $\left\{m_{d,i},\dots,m_{i,i}\right\} \subset \left\{m_{d,j},\dots,m_{j,j}\right\}$ for $1 \geq i \ge j \geq d$,
	\item $m_{i,i} \notin \left\{m_{i+1,i+1},\dots,m_{d,i+1}\right\}$ for $i=1,\dots,d-1$,
	\item for all $j=d-2,\dots,1$, $i=j+1,\dots,d$, there exist $\left(k,\ell\right)$ with $k < j$ and $\ell < k$ such that
	\begin{equation}\label{eq:proximitycondition3}
		\begin{aligned}
			& \left\{m_{i,j}, \left\{m_{d,j},\dots,m_{i+1,j}\right\}\right\} =
			 \left\{m_{k,k}, \left\{m_{1,k},\dots,m_{\ell,k}\right\}\right\} \mbox{ or } \\
			&\left\{m_{i,j}, \left\{m_{d,j},\dots,m_{i+1,j}\right\}\right\} =
			\left\{m_{\ell,k}, \left\{m_{1,k},\dots,m_{\ell-1,k},m_{k,k}\right\}\right\}.
		\end{aligned}
	\end{equation}
\end{enumerate}
The last property is reflecting the proximity condition. Conditions on $\bm{M}$ can be checked very quickly algorithmically.

\begin{example}[R-vine in 6 dimensions]\label{ex:exvine1}
	The R-vine matrix $M$ describes the R-vine in Figure \ref{fig:exvine1:1} as follows. Edges in $T_1$ are pairs of the main diagonal and the lowest row, e.\,g.\ $\left(2{,}1\right)$, $\left(6{,}2\right)$, $\left(3{,}6\right)$, etc. $T_2$ is given by the main diagonal and the second last row conditioned on the last row, e.\,g.\ $6{,}1|2$; $3{,}2|6$, etc. Higher order trees are encoded similarly.
	\begin{figure}[H]
		\centering
		\includegraphics[width=0.3\textwidth]{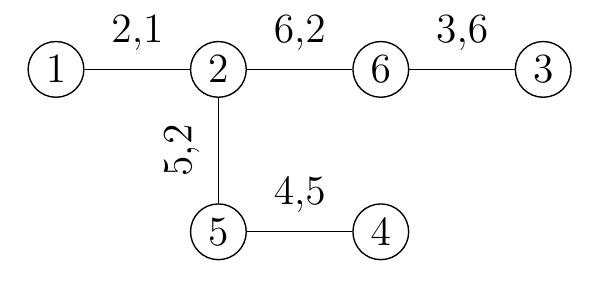}
		\includegraphics[width=0.3\textwidth]{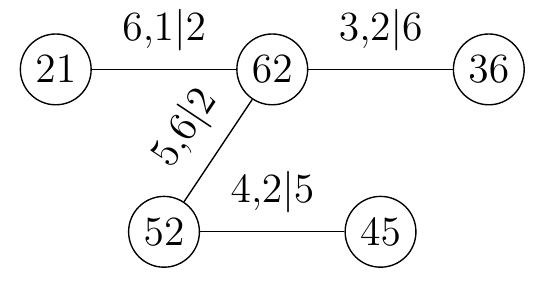}\\
		\includegraphics[width=0.3\textwidth]{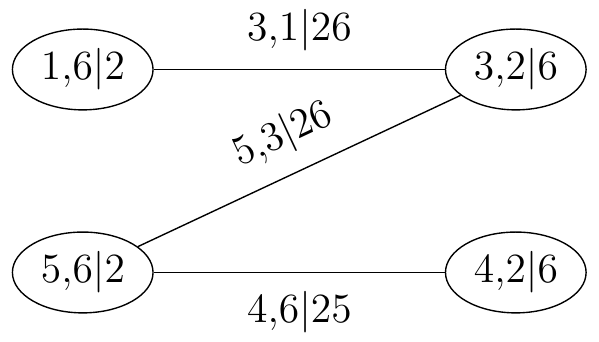}
		\includegraphics[width=0.2\textwidth]{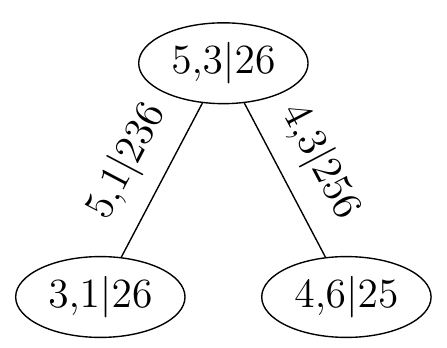}
		\includegraphics[width=0.25\textwidth]{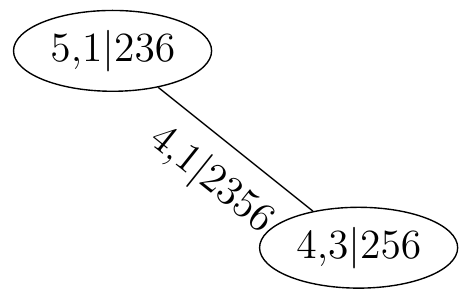}
		\caption{R-vine trees $T_1, T_2$ (top), $T_3, T_4, T_5$ (bottom), left to right.}
		\label{fig:exvine1:1}
		\vspace{-0.75cm}
	\end{figure}
	The associated R-vine matrix $M$ is given by
	\begin{equation*}
		M = 	\left(
		\begin{array}{cccccc}
			4&&&&&\\
			1&5&&&&\\
			3&1&3&&&\\
			6&3&1&6&&\\
			2&6&2&1&2&\\
			5&2&6&2&1&1
		\end{array}
		\right)
	\end{equation*}

	With $c_{j,\ell|D} \coloneqq c_{j,\ell;D}\left(F\left(x_i|\bm{x}_D\right),F\left(x_j|\bm{x}_D\right)\right)$ for conditioning vector $\bm{x}_D$, $\mathbf{x} = \left(x_1,\dots,x_{6}\right)$, $f_i \coloneqq f_i(x_i)$, the density becomes
	\begin{equation*}
		\begin{aligned}
			f\left(\mathbf{x}\right) = &f_1 \times f_2 \times f_3 \times f_4 \times f_5 \times f_6 \times c_{2,1} \times c_{6,2} \times c_{3,6} \times c_{5,2}\times c_{4,5} \times \\ & c_{6,1|2} \times c_{3,2|6}  \times c_{5,6|2} \times c_{4,2|5} \times c_{3,1|26} \times c_{5,3|26} \times c_{4,6|25} \times c_{5,1|236} \times c_{4,3|256} \times c_{4,1|2356}.
		\end{aligned}
	\end{equation*}
	The corresponding pair copula families and their parameters can also be stored in lower triangular family and parameter matrices $\bm{\Gamma} = \left(\gamma_{i,j}\right)_{i=1,\dots,d; j=1,\dots,d}$ and $\bm{P} = \left(p_{i,j}\right)_{i=1,\dots,d; j=1,\dots,d}$. Thus, the family and parameters of the pair copula $6,1|2$ described by $m_{5,4}$ are given by $\gamma_{5,4}$ and $p_{5,4}$. When two-parametric pair copulas are considered, an additional parameter matrix $\bm{P}_2$ is used similarly.
\end{example}
Since we are interested in high dimensional applications, model reduction plays an essential role. Overall, there are $d\left(d-1\right)/2$ edges, thus, model complexity increases quadratically in $d$. This can be simplified by only modeling the first $k$ trees and assuming (conditional) independence for the remaining higher $d-1-k$ trees, see \cite{brechmann-etal} for a discussion. If $k \in \left\{1,\ldots,d-2\right\}$, then a $k$-truncated R-vine is an R-vine where each pair copula density $c_{j\left(e\right),\ell\left(e\right);D\left(e\right)}$ assigned to an edge $e \in \left\{E_{k+1},\ldots,E_{d-1}\right\}$ is represented by the independence copula density $c^\perp\left(u_1,u_2\right) \equiv 1$. In a $k$-truncated R-vine, the second outer product in \eqref{eq:vinedensity} has $k$ instead of $d-1$ factors.
For an R-vine model with the parameter set $\bm{\Theta} =  \left(\VVV,\mathcal{B}\left(\VVV\right),\theta\left(\mathcal{B}\left(\VVV\right)\right)\right)$, consider $n$ replications of $d$ dimensional data $\left(\bm{x}_1,\dots,\bm{x}_n\right)^T \in \mathbb{R}^{n \times d}$ with $\bm{x}_i \in \mathbb{R}^d$ for $i=1,\dots,n$. We neglect the marginal distributions $f_i\left(x_i\right),\ i=1,\dots,d$, the log-Likelihood on the \textit{u-scale} is
\begin{equation*}
\LLL\left(\bm{\Theta},\left(\bm{x}_1,\dots,\bm{x}_n\right)\right)= 	\sum_{i=1}^n~\sum_{i=1}^{d-1}~\sum_{e\in E_i}~ \log\bigg(c_{j\left(e\right),\ell\left(e\right);D\left(e\right)}\Big( F\left(x_{j\left(e\right)}|\bm{x}_{D\left(e\right)}\right),\notag F\left(x_{\ell\left(e\right)}|\bm{x}_{D\left(e\right)}\right)\Big)\biggr).
\end{equation*}
Since the log-Likelihood always increases whenever more parameters enter the model, it is not advisable to use it for especially sparse structures since models will contain too many parameters of which a large portion do not contribute significantly to the model fit. Thus, there exist penalized goodness-of-fit measures which require that the log-Likelihood increases significantly to prefer a larger model. Such measures are the \textit{Akaike information criterion (AIC)} \cite{Akaike1973} and the \textit{Bayesian information criterion (BIC)} \cite{Schwarz1978}. Abbreviate $\LLL\left(\bm{\Theta}\right) \coloneqq \LLL\left(\bm{\Theta},\left(\bm{x}_1,\dots,\bm{x}_n\right)\right)$ and define
\begin{equation*}
	\begin{aligned}
		AIC\left(\bm{\Theta}\right) &= -2 \LLL\left(\bm{\Theta}\right) + 2p\\
		BIC\left(\bm{\Theta}\right) &= -2 \LLL\left(\bm{\Theta}\right) + \log\left(n\right)p,
	\end{aligned}
\end{equation*}
where $p$ equals the number of parameters in the model $\bm{\Theta}$. For $n \geq 8$, \textit{BIC} will penalize more than \textit{AIC}. If the number of possible parameters in an R-vine $q\left(d\right) = 2 \times d\left(d-1\right)/2$ is greater or equal than the sample size and the model is comparably small, BIC is no longer consistent and will penalize too little. For high dimensional data, this assumption is reasonable and we use a modified version of BIC (\textit{mBIC}) as in \cite{Frommletetal2011},
\begin{equation}\label{eq:mbic}
	mBIC\left(\bm{\Theta}\right) = -2 \LLL\left(\bm{\Theta}\right) + p\log\left(nq^2\right) - 2 \log\left(p!\right)- \sum_{j=1}^p~\log\left(\log\left(nq^2/j\right)\right).
\end{equation}

\section{Structural equation models (SEMs)}\label{sec:sem}
Our approach connects the R-vine structure to \textit{structural equation models (SEMs)}. For this, we utilize the approach of \cite{BrechmannJoe2014}, who give a representation of $k$-truncated Gaussian R-vines in terms of \textit{structural equation models (SEMs)}. SEMs are often used to model the influence of unobservable latent variables, see e.\,g.\ \cite{Kaplan2009}, \cite{Hoyle1995} or \cite{Bollen1989}. We want to stress that we are not considering latent variables in this paper and are thus only dealing with actual observations. Given a Gaussian R-vine with structure $\VVV$, we define a SEM corresponding to $\VVV$ denoted by $\SSS\left(\VVV\right)$. Let $\VVV = \left(T_1,\dots,T_{d-1}\right)$ be an R-vine tree sequence and assume without loss of generality $\left\{1,2\right\} \in T_1$. For $j = 3,\dots,d$ denote the edges in $T_1$ by $\left\{j, \kappa_1\left(j\right)\right\}$ using an \textit{assignment function} $\kappa_1\left(j\right)$, $j=2,\dots,d$. For higher trees, we generalize $\kappa_i$ for $i=2,\dots,i-1$. Thus, the trees $T_i$ contain edges $j,\kappa_i\left(j\right)|\kappa_1\left(j\right),\dots,\kappa_{i-1}\left(j\right) \in T_i$ for $i = 2,\dots,d-1$. Based on the R-vine structure $\VVV$, define $\SSS\left(\VVV\right)$ by
\begin{equation}\label{eq:sem}
	\begin{aligned}
		X_1 &= \psi_1 \epsilon_1,\\
		X_2 &= \varphi_{2,1} X_1 + \psi_2 \epsilon_2,\\
		X_j &= \sum_{i = 1}^{j-1}~\varphi_{j,\kappa_i\left(j\right)} X_{\kappa_i\left(j\right)} + \psi_j \epsilon_j,
	\end{aligned}
\end{equation}
with $\epsilon_j \sim \NNN\left(0,1\right)$ i.i.d. and $\psi_j$ such that $\Variance\left(X_j\right) = 1$ for $j = 1,\dots,d$.
\cite{BrechmannJoe2014} assume a $k$-truncated R-vine and restate the SEM in \eqref{eq:sem} with
\begin{equation*}
	\begin{aligned}
		X_j &= \sum_{i = 1}^{\max\left(j-1,k\right)}~\varphi_{j,\kappa_i\left(j\right)} X_{\kappa_i\left(j\right)} + \psi_j \epsilon_j,\ j = 3,\dots,d.
	\end{aligned}
\end{equation*}
Thus, we have for each edge $e \in E_i,\ i=k+1,\dots,d-1$ that for $j=3,\dots,d$:
\begin{equation}\label{eq:trvsparseidentity}
	c_{j\left(e\right),\kappa_i\left(j\left(e\right)\right);\kappa_1\left(j\left(e\right)\right),\dots,\kappa_{i-1}\left(j\left(e\right)\right)} \equiv 1 \Rightarrow \varphi_{j, \kappa_i\left(j\right)} = 0.
\end{equation}
The first step to generalize this implication is that we not only allow for a specific truncation level $k = 1,\dots,d-1$. Furthermore, we want to set specific regression coefficients $\varphi_{j, \kappa_i\left(j\right)}$ to zero, also for $i < k$. Additionally, we generalize the ordering of the equations from first to last using an ordering function $\eta: \left\{1,\dots,d\right\} \rightarrow \left\{1,\dots,d\right\}$. Thus, $X_{\eta\left(j\right)}$ is on the left hand side of the $j$-th equation and has at most $j$ right hand summands, including the error term, i.\,e.\ we obtain a triangular structure. We rewrite \eqref{eq:sem} as
\begin{equation}\label{eq:sem:eta}
	\begin{aligned}
		X_{\eta\left(1\right)} &={} \psi_{\eta\left(1\right)} \epsilon_{\eta\left(1\right)},\\
		X_{\eta\left(2\right)} &={} \varphi_{\eta\left(2\right),\kappa_1\left({\eta\left(2\right)}\right)} X_{\kappa_1\left({\eta\left(2\right)}\right)} + \psi_{\eta\left(2\right)} \epsilon_{\eta\left(2\right)},\\
		X_{\eta\left(j\right)} &={} \sum_{i = 1}^{j-1}~ \varphi_{\eta\left(j\right),\kappa_i\left({\eta\left(j\right)}\right)} X_{\kappa_i\left({\eta\left(j\right)}\right)} + \psi_{\eta\left(j\right)} \epsilon_{\eta\left(j\right)}.
	\end{aligned}
\end{equation}
We define some additional terminology to deal with zero regression coefficients.

\begin{Definition}[SEM regressor sets]\label{def:semregressorsets}
	Consider a SEM as in \eqref{eq:sem:eta} with ordering function $\eta$. Then, $X_{\eta\left(j\right)}$ has at most $j-1$ potential regressors $\kappa_i\left(\eta\left(j\right)\right)$ for $i=1,\dots,j-1$. We define the set of potential regressors of $X_{\eta\left(j\right)}$ by $\RRR\left(\eta\left(j\right)\right) = \left\{\eta\left(1\right),\dots,\eta\left(j-1\right)\right\}$, i.\,e.\ the left hand side indices of the previous $j-1$ structural equations. Define the set $\RRR_1\left(\eta\left(j\right)\right) = \left\{\kappa_i\left(\eta\left(j\right)\right),\ i = 1,\dots,j-1: \varphi_{\eta\left(j\right), \kappa_i\left(\eta\left(j\right)\right)} \neq 0 \right\}$, the set of actual regressors of $X_{\eta\left(j\right)}$. $\RRR_0\left(\eta\left(j\right)\right) = \RRR\left(\eta\left(j\right)\right) \setminus \RRR_1\left(\eta\left(j\right)\right)$ is the set of unused regressors.
\end{Definition}
We visualize the concepts in the following example. Recall that $j$ refers to the $j$-th row in the SEM and $\eta\left(j\right)$ to the corresponding left hand side index of the $j$-th row.

\begin{example}[Example \ref{ex:exvine1} cont.]\label{ex:vine1sem}
	Following our previous example, the R-vine matrix $\bm{M}$ gives rise to the following values of the ordering function $\eta$ and the assignment function $\kappa$. Considering $\eta$, we have the main diagonal $\diag\left(\bm{M}\right) = \left(4,5,3,6,2,1\right)=\left(\eta\left(d\right),\dots,\eta\left(1\right)\right)$, see also Table \ref{tab:ex1:etakappa}, left two columns. Since R-vine matrices are most often denoted as lower-diagonal matrices in the literature, we have $m_{1,1}=\eta\left(d\right),\dots,m_{d,d}=\eta\left(1\right)$. The values of the assignment function $\kappa$ can be read column-wise from $\bm{M}$. For example, consider $\bm{M}_{,j}$, the $j$-th column of $\bm{M}$ with $m_{j,j} = \eta\left(d-j+1\right)$. Then, $\kappa_1\left({\eta\left(d-j+1\right)}\right) = m_{d,j}$ and $\kappa_2\left({\eta\left(d-j+1\right)}\right) = m_{d-1,j}$. Generally, we obtain for $i=1,\dots,d-j$:
	\begin{equation*}
		\kappa_i\left({\eta\left(d-j+1\right)}\right) = m_{d-i+1,j},\ j=1,\dots,d-2.
	\end{equation*}
	The values of $\kappa$ can also be written in tabular form, see Table \ref{tab:ex1:etakappa}. The $i$-th row of this table corresponds to column $d-i+1$ of the R-vine matrix $\bm{M}$. For example, consider the first column of $\bm{M}$, i.\,e.\ $j=1$ with $m_{j,j}=\eta\left(6-1+1\right) = \eta\left(6\right) = 4$, according to Table \ref{tab:ex1:etakappa}. Correspondingly $\kappa_1\left({\eta\left(6-1+1\right)}\right) = \kappa_1\left({\eta\left(6\right)}\right) = \kappa_1\left(4\right) = 5 = m_{6,1}$, see also Table \ref{tab:ex1:etakappa}. 
	\begin{table*}
		\centering
		\begin{tabular}{c||c|ccccc}
			\hline
			$j$ & $\eta\left(j\right)$ & $\kappa_1\left(\eta\left(j\right)\right)$ & $\kappa_2\left(\eta\left(j\right)\right)$ & $\kappa_3\left(\eta\left(j\right)\right)$ & $\kappa_4\left(\eta\left(j\right)\right)$ & $\kappa_5\left(\eta\left(j\right)\right)$\\
			\hline \hline
			1 & 1 & - & - & - & - & - \\
			2 & 2 & $1=m_{6,5}$ & - & - & - & - \\
			3 &	6 & $2=m_{6,4}$ & $1=m_{5,4}$ & - & - & - \\
			4 &	3 & $6=m_{6,3}$ & $2=m_{5,3}$ & $1=m_{4,3}$ & - &- \\
			5 &	5 & $2=m_{6,2}$ & $6=m_{5,2}$ & $3=m_{4,2}$ & $1=m_{3,2}$ & -\\
			6 &	4 & $5=m_{6,1}$ & $2=m_{5,1}$ & $6=m_{4,1}$ & $3=m_{3,1}$ & $1=m_{2,1}$ \\
			\hline
		\end{tabular}
		\caption{Example \ref{ex:exvine1}: Inverse of ordering function $\eta$ and assignment function $\kappa$}
		\label{tab:ex1:etakappa}
	\end{table*}
The R-vine Matrix $\bm{M}$ is given by
	\begin{equation*}
		\left(
		\begin{array}{cccccc}
			m_{1,1}&&&&&\\
			m_{2,1}&m_{2,2}&&&&\\
			m_{3,1}&m_{3,2}&m_{3,3}&&&\\
			m_{4,1}&m_{4,2}&m_{4,3}&m_{4,4}&&\\
			m_{5,1}&m_{5,2}&m_{5,3}&m_{5,4}&m_{5,5}&\\
			m_{6,1}&m_{6,2}&m_{6,3}&m_{6,4}&m_{6,5}&m_{6,6}
		\end{array}
		\right)
		=
		\left(
		\begin{array}{cccccc}
			4&&&&&\\
			1&5&&&&\\
			3&1&3&&&\\
			6&3&1&6&&\\
			2&6&2&1&2&\\
			5&2&6&2&1&1
		\end{array}
		\right)
	\end{equation*}
	
	We now want to evaluate the correspondence between independence copulas in the R-vine and zero coefficients in the SEM. Assume the following lower triangular family matrix $\bm{\Gamma} = \left(\gamma_{i,j}\right)_{i=1,\dots,d; j=1,\dots,d}$ with $0$ representing independence and $1$ indicating a Gaussian copula. 
	\newline
	\begin{equation*}
		\bm{\Gamma} = 
		\left(
		\begin{array}{cccccc}
			-&&&&&\\
			\gamma_{2,1}&-&&&&\\
			\gamma_{3,1}&\gamma_{3,2}&-&&&\\
			\gamma_{4,1}&\gamma_{4,2}&\gamma_{4,3}&-&&\\
			\gamma_{5,1}&\gamma_{5,2}&\gamma_{5,3}&\gamma_{5,4}&-&\\
			\gamma_{6,1}&\gamma_{6,2}&\gamma_{6,3}&\gamma_{6,4}&\gamma_{6,5}&-
		\end{array}
		\right)
		=\left(
		\begin{array}{cccccc}
			-&&&&&\\
			0&-&&&&\\
			1&1&-&&&\\
			0&0&1&-&&\\
			1&0&0&1&-&\\
			1&1&1&1&1&-
		\end{array}
		\right)
	\end{equation*}
	
	The zeros in the family matrix $\bm{\Gamma}$, i.\,e.\ independence copulas, are reflected by zero coefficients in the SEM. For $j=1,\dots,d-2$ and $i=1,\dots,d-j$ we have
	\begin{equation*}
		\gamma_{d-i+1,j} = 0 \Rightarrow \varphi_{\eta\left(d-j+1\right),\kappa_i\left(\eta\left(d-j+1\right)\right)} = 0.
	\end{equation*}
	We emphasize that only the parameter value $\varphi$ is set to zero. The assignment function $\kappa$ is unchanged since it is necessary to determine a valid R-vine structure. This way, we impose independence, i.\,e.\ sparsity in the R-vine which is reflected by the corresponding SEM. We now illustrate how this choice affects $\RRR$, $\RRR_0$ and $\RRR_1$.
	\begin{equation}\label{eq:semeqexample}
		\begin{aligned}
			X_1 &= \psi_1 \epsilon_1,\\
			X_2 &= \varphi_{2,1} X_1 + \psi_2 \epsilon_2,\\
			X_6 &= \varphi_{6,2} X_2 + \varphi_{6,1} X_1 + \psi_6 \epsilon_6,\\
			X_3 &= \varphi_{3,6} X_6 + \varphi_{3,1} X_1 + \psi_3 \epsilon_3,\\
			X_5 &= \varphi_{5,2} X_2 + \varphi_{5,1} X_1 + \psi_5 \epsilon_5,\\
			X_4 &= \varphi_{4,5} X_5 + \varphi_{4,2} X_2 + \varphi_{4,3} X_3 + \psi_3 \epsilon_3.
		\end{aligned}
	\end{equation}
	\begin{table}[H]
		\centering
		\begin{tabular}{c|ccc}
			\hline
			$\eta\left(j\right)$ & $\RRR(\eta\left(j\right))$ & $\RRR_1(\eta\left(j\right))$ & $\RRR_0(\eta\left(j\right))$ \\
			\hline \hline
			$1$ & $\emptyset$ 			& $\emptyset$ 	& $\emptyset$ \\
			$2$ & $\left\{1\right\}$ 			& $\left\{1\right\}$ 	& $\emptyset$ \\
			$6$ & $\left\{2,1\right\}$ 			& $\left\{2,1\right\}$  & $\emptyset$ \\
			$3$ & $\left\{6,2,1\right\}$ 		& $\left\{6,1\right\}$	& $\left\{2\right\}$ \\
			$5$ & $\left\{2,6,3,1\right\}$ 		& $\left\{2,1\right\}$ 	& $\left\{6,3\right\}$\\
			$4$ & $\left\{5,2,6,3,1\right\}$ 	& $\left\{5,2,3\right\}$ & $\left\{6,1\right\}$ \\
			\hline
		\end{tabular}
		\caption{Example \ref{ex:exvine1}: Sets $\RRR$, $\RRR_1$, $\RRR_0$}
		\label{tab:ex1:rsets}
	\end{table}
	In other words, the non-zero coefficients in the SEM \eqref{eq:semeqexample} are drawn from the corresponding columns of the R-vine structure matrix $\bm{M}$ where the family matrix $\bm{\Gamma}$ is non-zero. Consider an arbitrary column $j = 1,\dots,5$ in the matrix $\bm{M}$. The non-zero entries $\left(\gamma_{d,j},\dots,\gamma_{j+1,j}\right)$ correspond to $\RRR_1\left(\eta\left(d-j+1\right)\right)$. For example, if we consider again the first column of $\bm{M}$, $\bm{M}_{,j}$ for $j=1$  and $\left(\gamma_{6,1},\dots,\gamma_{2,1}\right) = \left(1,1,0,1,0\right)$. Using this vector to obtain the non-zero entries from the R-vine structure matrix $\bm{M}$, we have the first column $\bm{M}_{d:2,1} = \left(5,2,6,3,1\right)$ and thus the non-zero entries $\left(5,2,3\right)$ as in \eqref{eq:semeqexample} and Table \ref{tab:ex1:rsets} for $\RRR_1\left(\eta\left(d-1+1\right)\right) = \RRR_1\left(\eta\left(6\right)\right) = \RRR_1\left(4\right)$.
\end{example}
Having characterized the connection between R-vines and SEMs, our goal is now to find an inverse transformation. More precisely, given high dimensional data, we want estimate a SEM where many of the coefficients are zero. For simplicity, assume $\eta\left(j\right)=j$ for $j=1,\dots,d$. For each structural equation, we obtain a set $\RRR_0\left(j\right)$ with $\left|\RRR_0\left(j\right)\right| > 0$. This leaves us with a sparse SEM as in \eqref{eq:sem},
\begin{equation}\label{eq:sem:sparse}
	\begin{aligned}
		X_1 &= \psi_1 \epsilon_1,\\
		X_2 &= \varphi_{2,1} X_1 + \psi_2 \epsilon_2,\\
		X_j &= \sum_{i \in \RRR_1\left(\eta\left(j\right)\right)}~\varphi_{j,\kappa_i\left(j\right)} X_{\kappa_i\left(j\right)} + \psi_j \epsilon_j,\ j=3,\dots,d.
	\end{aligned}
\end{equation}
Under additional assumptions, this SEM can also be written as an R-vine with structure matrix $\widehat{\bm{M}}$ and family matrix $\widehat{\bm{\Gamma}}$. Because of the zero-coefficients in $\RRR_0\left(j\right)$, entries in the family matrix $\widehat{\bm{\Gamma}}$ can be set to $0$, i.\,e.\ representing the independence copula. This means, we want to generalize the implication \eqref{eq:trvsparseidentity} in such a way that we have for each edge $e \in E_i,\ i=1,\dots,d-1$ and $j=3,\dots,d$:
\begin{equation}\label{eq:intuitionsemrepresentation}
	\varphi_{j, \kappa_i\left(j\right)} = 0  \Rightarrow  c_{j\left(e\right),\kappa_i\left(j\left(e\right)\right);\kappa_1\left(j\left(e\right)\right),\dots,\kappa_{i-1}\left(j\left(e\right)\right)} = 1.
\end{equation}
Thus, we obtain a sparse R-vine model. This model is not restricted to a joint Gaussian probability distribution as our SEM is. We can estimate the marginal distributions entirely independent of the dependence behaviour and use vast numbers of parametric and non-parametric pair copulas to describe the joint distribution. To describe more precisely what is motivated by \eqref{eq:intuitionsemrepresentation}, we now introduce an R-vine representation of SEM.
\begin{Definition}[R-vine representation of a SEM]\label{def:rvinerepsem}
	Consider a SEM in $d$ dimensions, where we assume without loss of generality $\eta\left(j\right) \equiv j$ for $j=1,\dots,d$.
	\begin{equation}\label{eq:sem:def}
		\begin{aligned}
			X_1 &= \psi_1 \epsilon_1,\\
			X_2 &= \varphi_{2,1} X_1 + \psi_2 \epsilon_2,\\
			X_j &= \sum_{i = 1}^{j-1}~\varphi_{j,\kappa_i\left(j\right)} X_{\kappa_i\left(j\right)} + \psi_j \epsilon_j.
		\end{aligned}
	\end{equation}
	The SEM \eqref{eq:sem:def} has an R-vine representation $\VVV$ if there exists an R-vine tree sequence $\VVV=\left(T_1,\dots,T_{d-1}\right)$ such that for $j=2,\dots,d$ and $i = 1,\dots,j-1$ we have
	\begin{align*}
		j{,}\kappa_i\left(j\right)|\kappa_1\left(j\right),\dots,\kappa_{j-1}\left(i\right) \in T_j.
	\end{align*}
\end{Definition}
To put it in a nutshell, the $j$-row of the SEM corresponds to column $d-j+1$ of the R-vine matrix for $j=1,\dots,d$. This definition connects SEMs and R-vines. Based on this, we can consider setting specific regressors in the SEM to zero to obtain a sparse R-vine model. We note two caveats of this approach. First, of all, not every SEM with specific coefficients set to zero reflects a R-vine structure, since the proximity condition has to hold for the R-vine structure. Second, a SEM does not necessarily determine the R-vine structure uniquely. We give examples for these assertions and move on to sketch the general approach.

\begin{example}[SEM without R-vine representation]\label{ex:semworvine}
	Consider the following SEM in $5$ dimensions.
	\begin{equation*}
		\begin{aligned}
			X_1 &= \psi_1 \epsilon_1\\
			X_2 &= \varphi_{2,1} X_1 + \psi_2 \epsilon_2\\
			X_3 &= \varphi_{3,1} X_1 + \varphi_{3,2} X_2 + \psi_3 \epsilon_3\\
			X_4 &= \varphi_{4,1} X_1 + \varphi_{4,2} X_2 + \psi_4 \epsilon_4\\
			X_5 &= \varphi_{5,3} X_3 + \varphi_{5,4} X_4 + \psi_5 \epsilon_5
		\end{aligned}
	\end{equation*}
	If we now want to find a representing R-vine structure, the R-vine trees $T_1$ and $T_2$ must have edges in terms of the assignment function $\kappa$ as we saw from definition \ref{def:rvinerepsem}. Since we have at most two right hand side summands, we need to find values for $\kappa_i\left(j\right)$ for $j = 3,4,5$ and $i=1,2$ such that the following holds:
	\begin{equation*}
		\begin{aligned}
			& i{,}\kappa_1\left(i\right) & \in T_1,\\
			& i{,}\kappa_2\left(i\right)|\kappa_1\left(i\right) & \in T_2.
		\end{aligned}
	\end{equation*}
	Assume without loss of generality the following edges are chosen in the first tree $T_1$: $ \left\{\left\{2{,}\kappa_1\left(2\right)\right\},\left\{3{,}\kappa_1\left(3\right)\right\},\left\{4{,}\kappa_1\left(4\right)\right\},\left\{5{,}\kappa_1\left(5\right)\right\}\right\} = \left\{\left\{2{,}1\right\},\left\{3{,}1\right\},\left\{4{,}1\right\},\left\{5{,}3\right\}\right\} \in E_1$.
	Now, we can not set $\kappa_2\left(5\right)= 4$ to obtain $5{,}4|3 \in T_2$ as required. This is since $5{,}4|3 = \left\{\left\{5,3\right\},\left\{4,3\right\}\right\}$, but $\left\{3{,}4\right\} \notin T_1$. Note additionally that we can not have more than four edges in $T_1$, since otherwise, it would not be a tree.
\end{example}
Next, we show an example of that two R-vines with identical SEM representations.
\begin{example}[Different $2$-truncated R-vines with identical SEM representation in $4$ dimensions]\label{ex:semunique}
	Consider the following two $2$-truncated R-vines and their SEM representations.\\
	\begin{figure}[H]
		\centering
		\includegraphics[width=0.12\textwidth, trim={0.1cm 0.1cm 0.1cm 0.1cm},clip]{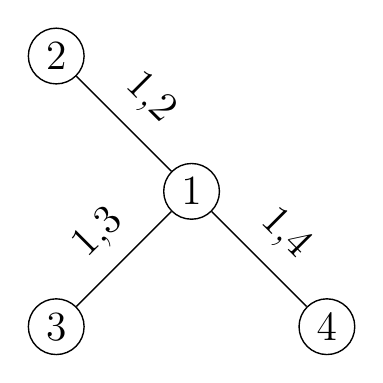}
		\includegraphics[width=0.12\textwidth, trim={0.1cm 0.1cm 0.1cm 0.1cm},clip]{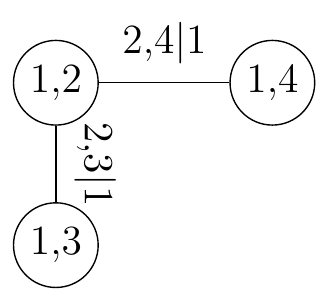}
		\caption{R-vine structure $\VVV_1$, corresponding to \eqref{eq:ex2:sem1}}
		\label{fig:exsemunique:vine1}
	\end{figure}
	\begin{figure}[H]
		\centering
		\includegraphics[width=0.12\textwidth, trim={0.1cm 0.1cm 0.1cm 0.1cm},clip]{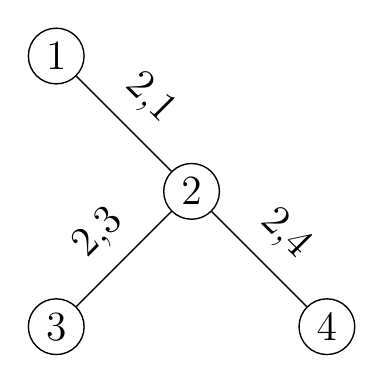}
		\includegraphics[width=0.12\textwidth, trim={0.1cm 0.1cm 0.1cm 0.1cm},clip]{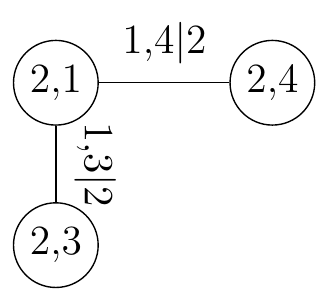}\\
		\caption{R-vine structure $\VVV_1$, corresponding to \eqref{eq:ex2:sem2}}
		\label{fig:exsemunique:vine2}
	\end{figure}
	Both have identical SEM representations, i.\,e.\ only looking at the corresponding equations without knowing exactly the assignment function $\kappa$ and thus, which regressor belongs to which R-vine tree, we are not able to distinguish between those two SEMs.\\
	\begin{equation}\label{eq:ex2:sem1}
		\begin{aligned}
			X_1 &= \psi_1 \epsilon_1,\\
			X_2 &= \varphi_{2,1} X_1 + \psi_2 \epsilon_2,\\
			X_3 &= \varphi_{3,1} X_1 + \varphi_{3,2} X_2 + \psi_3 \epsilon_3,\\
			X_4 &= \varphi_{4,1} X_1 + \varphi_{4,2} X_2 + \psi_4 \epsilon_4.
		\end{aligned}
	\end{equation}
	
	\begin{equation}\label{eq:ex2:sem2}
		\begin{aligned}
			X_1 &= \psi_1 \epsilon_1,\\
			X_2 &= \varphi_{2,1} X_1 + \psi_2 \epsilon_2,\\
			X_3 &= \varphi_{3,1} X_2 + \varphi_{3,2} X_1 + \psi_3 \epsilon_3,\\
			X_4 &= \varphi_{4,1} X_2 + \varphi_{4,2} X_1 + \psi_4 \epsilon_4.
		\end{aligned}
	\end{equation}
\end{example}
We will develop an approach which overcomes the restrictions sketched in the Examples \ref{ex:semunique} and \ref{ex:semworvine}. First, we will need to determine the R-vine structure based on the assignment function $\kappa$ before we consider the sets of zero-coefficients. The method we are going to use for this is the Lasso, which we will recapture briefly.

\section{The Lasso in linear regression}\label{sec:lasso}
In the most general case, consider a sample of $n$ observations $\left\{\mathbf{x}_i,y_i\right\}$, $i=1,\dots,n$, where $\mathbf{x}_i = \left(x_{i1},\dots,x_{ip}\right) \in \mathbb{R}^p$. We want to approximate $y_i$ given a set of linear predictors $x_{i,j}$
\begin{equation*}
	y_i = \varphi_0 + \sum_{j=1}^p~\varphi_j x_{i,j},
\end{equation*}
with unknown regression coefficients $\varphi_0$ and $\bm{\varphi} = \left(\varphi_1,\dots,\varphi_p\right)$. This is most often solved by minimizing the quadratic error with respect to $\varphi_0$ and $\bm{\varphi} = \left(\varphi_1,\dots,\varphi_p\right)$:
\begin{equation}\label{eq:leastsquares}
	\min_{\left(\varphi_0, \bm{\varphi}\right) \in \mathbb{R}^{p+1}}~\Bigg(\frac{1}{2n}\sum_{i=1}^n~\bigg(y_i - \varphi_0 - \sum_{j=1}^p~\varphi_j x_{i,j}\bigg)^2\Bigg).
\end{equation}
The solution to this optimization problem often contains many coefficients $\varphi_j \neq 0,\ j=1,\dots,p$. Thus, for $p$ large, the model becomes overly parametrized and hard to interpret. Yet, solving \eqref{eq:leastsquares} under the additional constraint 
\begin{equation}\label{eq:Lassot}
	\sum_{j=1}^p~\left|\varphi_j\right| \leq t,\ t \geq 0,
\end{equation}
yields a parsimonious model. This \textit{regularization technique} is called the Lasso and since its invention, see \cite{Tibshirani94regressionshrinkage}, proved very useful in many applications. By shrinking coefficients exactly to zero, it combines both parameter estimation and model selection in one step. It also works in cases where $p > n$, which are hard to solve otherwise. The Lasso is hence the method of choice when dealing with many possible predictors, of which only some contribute significantly to the model fit. For convenience, we will consider the following Lagrangian form of the optimization problem \eqref{eq:leastsquaresLasso}, which is equivalent to \eqref{eq:leastsquares} under the constraint \eqref{eq:Lassot}:
\begin{equation}\label{eq:leastsquaresLasso}
	\min_{\left(\varphi_0, \bm{\varphi}\right) \in \mathbb{R}^{p+1}}~\Bigg(\frac{1}{2n}\sum_{i=1}^n~\bigg(y_i - \varphi_0 - \sum_{\ell=1}^p~\varphi_\ell x_{i,\ell}\bigg)^2 + \lambda \sum_{\ell=1}^p \left|\varphi_\ell\right|\Bigg),
\end{equation}
for some $\lambda \geq 0$. One can show that a solution $\left(\widehat{\varphi}_0^\lambda,\widehat{\bm{\varphi}}_\lambda\right)$ of \eqref{eq:leastsquaresLasso} minimizes the problem in \eqref{eq:leastsquares} under the condition \eqref{eq:Lassot} with $t = \left|\widehat{\bm{\varphi}}_\lambda\right| = \sum_{j=1}^p~\left|\widehat{\varphi}_j^\lambda\right|$, see \cite{HastieTibshiraniWainwright2015}. We do not include an intercept in our considerations and thus set $\varphi_0 \equiv 0$ for the remainder of the paper. If we consider the problem \eqref{eq:leastsquaresLasso} and set $\lambda = \infty$, all coefficients $\widehat{\varphi}_j$, $j=1,\dots,p$ will be set to zero because of the penalization. Decreasing $\lambda > 0$, more and more coefficients become non-zero. This relationship between $\lambda > 0$ and $\widehat{\varphi}^\lambda_j,\ j=1,\dots,p$ is called the \textit{regularization path}. We formalize it by a set $\Lambda\left(\lambda\right)$ such that
\begin{equation*}
	\Lambda\left(\lambda\right) = \left\{\ell: \widehat{\varphi}_{\ell}^\lambda \neq 0 \mbox{ in } \widehat{\bm{\varphi}}_\lambda\right\}.
\end{equation*}
Thus, for each $\lambda > 0$ we are given the non-zero regression coefficients. How to choose $\lambda > 0$ is not obvious. Most often, \textit{k-fold cross-validation} is employed. Since it is not vital for the remainder of the paper, we describe it in Appendix \ref{sec:app:cv} and conclude with a brief example, introducing the concept of regularization paths.

\begin{example}[Lasso, regularization path, cross validation]\label{ex:lasso}
	We use the \texttt{worldindices} dataset, included in the \texttt{CDVine} package, see \cite{cdvine2013} comprising $d=6$ variables with $n=396$ observations on the \textit{u-scale}. More precisely, these are the stocks indices \verb|^|GSPC, \verb|^|N225, \verb|^|SSEC, \verb|^|GDAXI, \verb|^|FCHI, \verb|^|FTSE of the US, Japanese, Chinese, German, French and British stock markets. We transform our observations to the \textit{z-scale} using the normal quantile function, see page \pageref{en:scales} and denote them by $Z_i$, $i=1,\dots,6$ where $Z_1 \equiv \verb|^|GSPC$, $Z_2 \equiv \verb|^|N225$, and so on. Let us assume that we want to model the index $Z_3 = \verb|^|SSEC$ by the regressors \verb|^|GSPC, \verb|^|GDAXI, \verb|^|FCHI, \verb|^|FTSE, $Z_1,\ Z_4,\ Z_5,\ Z_6$, respectively. We write the regression equation
	\begin{equation*}
		Z_{i,3} = \varphi_0 + \sum_{j=1,4,5,6}~\varphi_j Z_{i,j},\ i=1,\dots,n,
	\end{equation*}
	with unknown regression coefficients $\varphi_0$ and $\bm{\varphi} = \left(\varphi_1,\dots,\varphi_p\right)$. We set $\varphi_0 \equiv 0$ and want to solve the regression problem with the Lasso. Thus we obtain the optimization problem
	\begin{equation}
		\min_{\bm{\varphi} \in \mathbb{R}^{4}}~\Biggl(\frac{1}{2n}\sum_{i=1}^n~\biggl(Z_{i,3} - \sum_{\ell=1,4,5,6}~\varphi_\ell Z_{i,\ell}\biggr)^2 + \lambda \sum_{\ell=1,4,5,6} \left|\varphi_\ell\right|\Biggr).
	\end{equation}
	The solution to this optimization problem is a regularization path, either along $\lambda > 0$ or $\sum_{i=1,4,5,6}~\left|\widehat{\varphi}_\ell\right|$, i.\,e.\ the $L_1$ norm of the regression vector. We use the R-package \texttt{glmnet} \citep{glmnet} to calculate the regularization paths with respect to the $L_1$ norm, see Figure \ref{fig:exlasso:regpatht} and $\log\left(\lambda\right)$..\\
	\vspace{-1cm}
	\begin{figure}[H]
		\centering
		\includegraphics[width=0.45\textwidth, trim={0.1cm 0.1cm 0.1cm 0.1cm},clip]{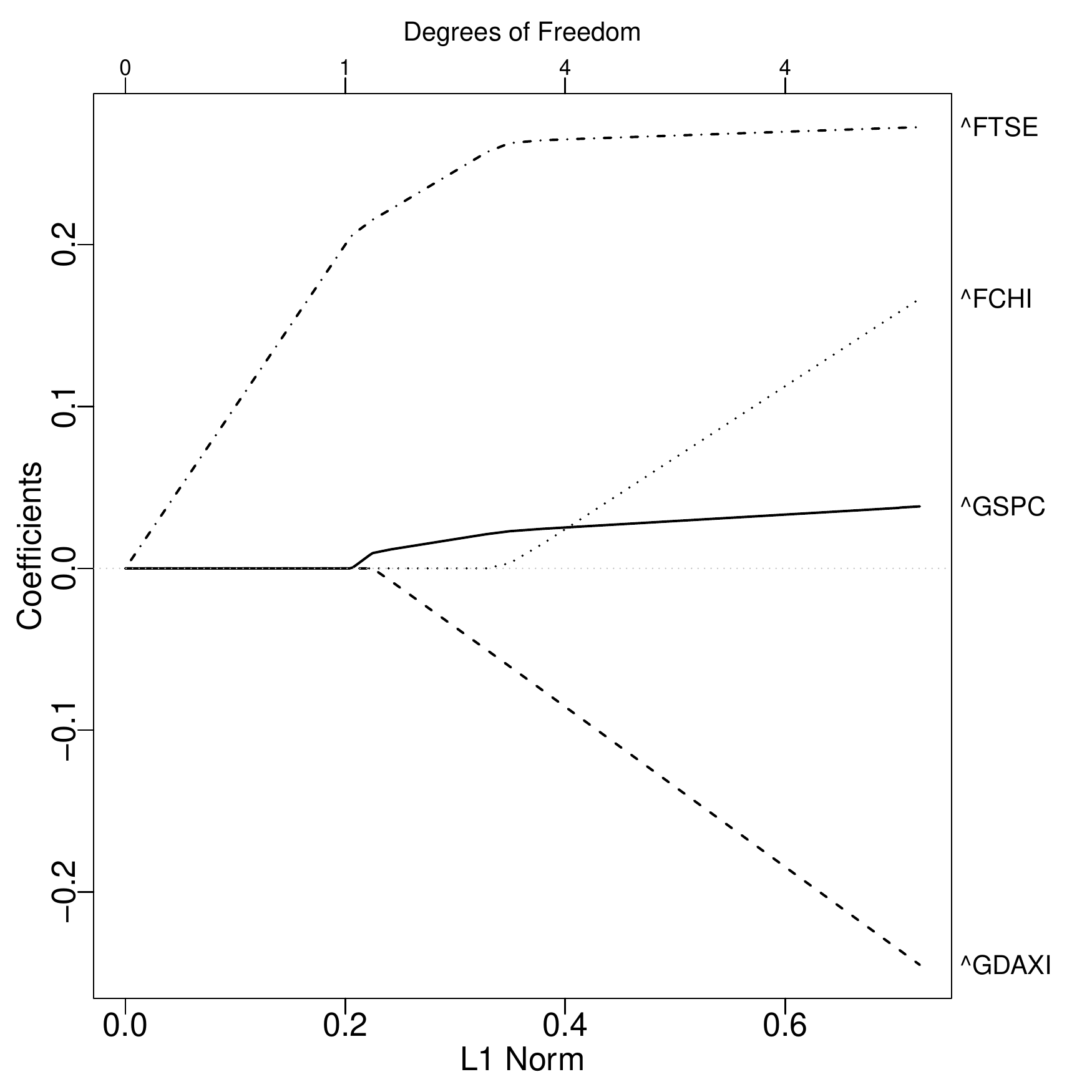}
		\includegraphics[width=0.45\textwidth, trim={0.1cm 0.1cm 0.1cm 0.1cm},clip]{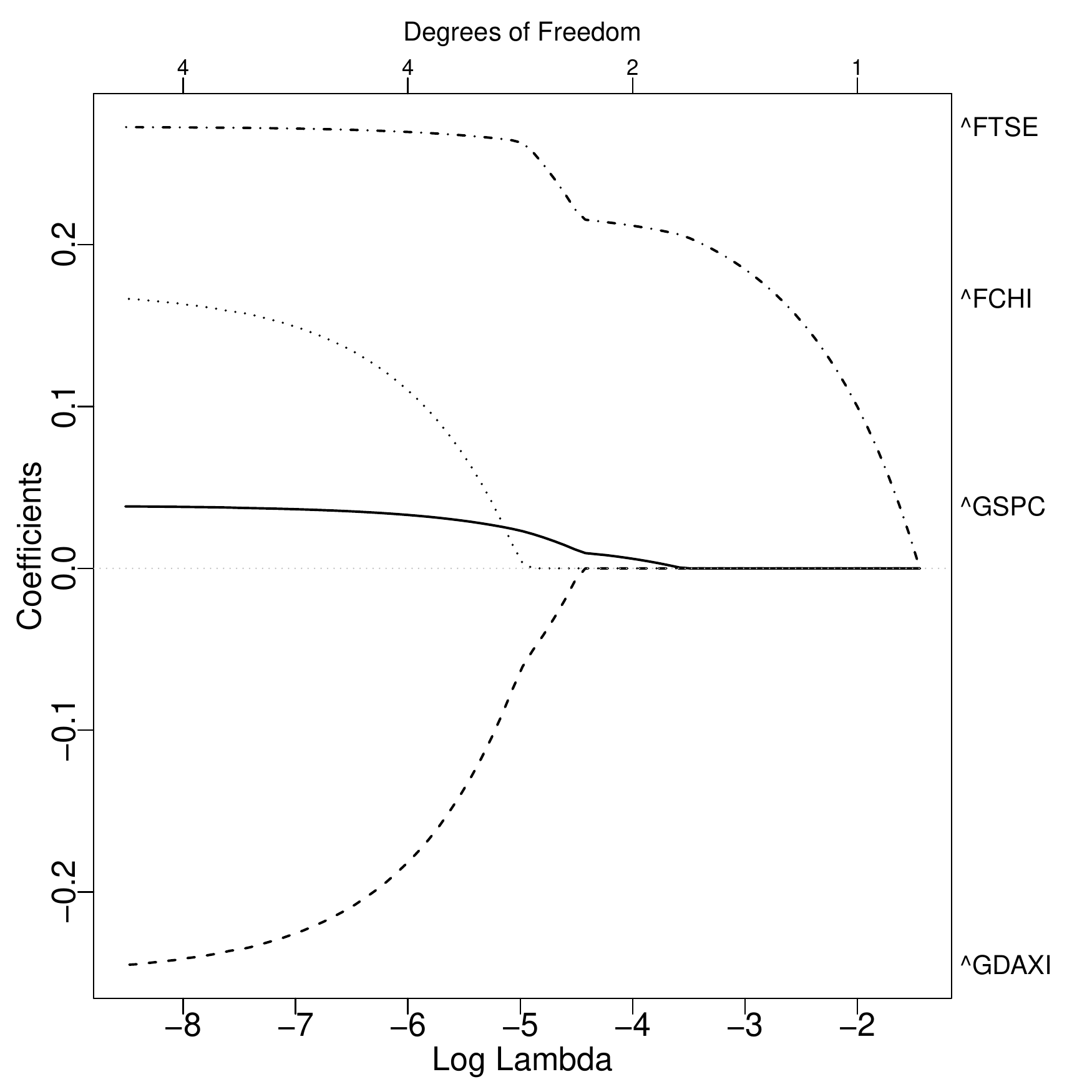}
		\caption{Regularization path of with respect to the $L_1$ norm of coefficients (left) and $\log\left(\lambda\right)$}
		\label{fig:exlasso:regpatht}
	\end{figure}
	\vspace{-0.5cm}
	We see that $Z_6 \equiv \verb|^|FTSE$ is the first non-zero coefficient along the regularization path. Additionally, we obtain that coefficients can of course also be negative and the regularization paths of different regressors may intersect. We denote the path by $\Lambda\left(0\right) = \left\{6,1,4,5\right\}$. Note that $\log\left(\lambda\right) < 0$ must not necessarily be the case as in this example. Above the plot, the corresponding number of non-zero parameters is indicated.
\end{example}
\section{Vine Copula structure selection with the Lasso}\label{sec:structureselectionLasso}
To use SEMs and the Lasso to calculate a vine copula structure, we proceed in three steps. First, we calculate an ordering function $\eta\left(i\right)$ for $i=1,\dots,d$ for the SEM ordering. Secondly, we identify the assignment function $\kappa_i$. Finally, we use the Lasso to identify the non-zero coefficient sets $\RRR_1\left(\eta\left(j\right)\right)$ for $j=1,\dots,d$. Before we calculate the ordering function $\eta$, recall the three different scales, \textit{x-scale}, \textit{u-scale} and \textit{z-scale}. Normally, data $\left(\bm{x}_1,\dots,\bm{x}_n\right)^T \in \mathbb{R}^{n \times d}$ with $\bm{x}_1 = \left(x_{11},\dots,x_{1d}\right)$ is obtained on the \textit{x-scale}. The transformation to the \textit{u-scale} is important for copula modeling as the marginal effects have then been removed from the data. The transformation to the \textit{z-scale} again is important for performing explorative data analysis. For example, considering contour shapes of bivariate data on the \textit{u-scale} is hard. However, on the \textit{z-scale}, deviations from normal dependence can be seen quite easily. Another advantage of the \textit{z-scale} over the \textit{x-scale} is that almost all data points will lie in an interval $\left[-3,3\right]$. Thus, performing regressions on such data will have standardized coefficients which eases the interpretation.
\subsection{Calculation of the ordering function}\label{subsec:calculationeta}
Assume for the moment we already have found an ordering $\eta$ and that it coincides with the ordering of the variables, i.\,e.\ $\eta\left(j\right) = j$ for $j=1,\dots,d$. In a SEM in the form of \eqref{eq:sem}, $X_j$ can have regressors $X_i$ for $i < j$, based on our model assumption. Thus, if we compute solutions for the $d$ equations
\begin{equation*}
	X_j = \sum_{i=1,i \neq j}^d~\beta_{i,j} X_i + \psi_j \epsilon_j,\ j=1,\dots,d,
\end{equation*}
we end up with a list of regression coefficients for each $X_j,\ j=1,\dots,d$. Moreover, if we solve these equations with the Lasso and some suitably chosen $\lambda \ge 0$, specific regression coefficients are set to zero. Considering all equations, some $X_i$ will occur more often with non-zero coefficients than others. Based on the SEM structure we have, it is beneficial to assign the regressors which occur often a low value of the ordering function $\eta$. In a SEM with such a structure, these $X_i$ which occurred often as regressors can then be chosen as regressors by the assignment function $\kappa$.
\begin{Definition}[Lasso Ordering]\label{def:Lassoordering}
	Consider $n$ samples from $\mathbf{X} =\left(X_1,\dots,X_d\right) \in \mathbb{R}^d$ and let $B \in \mathbb{R}^{\left(d-1\right)\times d}$ with columns $\bm{\beta}_j$, $j=1,\dots,d$ such that $\bm{\beta}_j = \left(\beta_{j,1},\dots,\beta_{j,j-1}, \beta_{j,j+1},\dots,\beta_{j,d}\right)$ are the Lasso solutions to the $d$ minimization problems 
	\begin{equation*}
		\min_{\bm{\beta}_j \in \mathbb{R}^{d-1}}~\Bigg(\frac{1}{2n}\sum_{j=1}^n~\bigg(x_{j} - \sum_{\ell=1, \ell \neq j}^d~\beta_{j,\ell} x_\ell\bigg)^2 + \lambda_\ell \sum_{\ell=1, \ell\neq j}^d~\left|\beta_{j,\ell}\right| \Bigg). 
	\end{equation*}
	For each possible regressor $j=1,\dots,d$, calculate the number of $\beta_{j,\ell}=0$ over all $\ell$ and assign the ones with highest occurrence the lowest number in the ordering function $\eta_L$. More precisely,
	\begin{equation*}
		\sum_{\ell = 1}^d~\mathds{1}_{\left\{\beta_{\eta_L\left(1\right),\ell}\neq 0\right\}} \leq \dots \leq \sum_{\ell = 1}^d~\mathds{1}_{\left\{\beta_{\eta_L\left(d\right),\ell}\neq 0\right\}}
	\end{equation*}
	The corresponding $\lambda_\ell$ are calculated via $k$-fold cross-validation. In case of ties, i.\,e.\ two or more variables are occurring equally often as regressors for the remaining variables, we choose the ordering of these variables randomly.
\end{Definition}
The intuition is similar to a method proposed by \cite{meinshausen2006} to find undirected graphical models. They use the Lasso to find neighbourhoods of nodes which are exactly the non-zero coefficient regressors calculated by the Lasso. We give a brief numerical example.

\begin{example}[Calculation of ordering function $\eta$]\label{ex:eta}
	We consider the \texttt{worldindices} dataset, included in the \texttt{CDVine} package, see \cite{cdvine2013} comprising $d=6$ variables with $n=396$ observations on the \textit{u-scale}. We transform our observations to the \textit{z-scale} using the normal quantile function.
	We calculate Lasso regression coefficients of $Z_j$ on $Z_{-j}$ for $j=1,\dots,6$. Of course, the number of non-zero regression coefficients depends on the choice of the penalization coefficient $\lambda_j$ for each regression on $Z_j$. Our experiments showed that it is feasible to choose $\lambda_j$ according to $k$-fold cross-validation.
	\begin{table}[H]
		\centering
		\begin{tabular}{lccc||r}
			\hline
			variable & id $j$ & \# occurrence & $\eta_L\left(j\right)$ & $\lambda_j$\\ 
			\hline
			\verb|^|GSPC & 1 & \textbf{2} & 4 & 0.170\\ 
			\verb|^|N225 & 2 & \textbf{2} & 5 & 0.129\\ 
			\verb|^|SSEC & 3 & \textbf{1} & 6 & 0.171\\ 
			\verb|^|GDAXI & 4 & \textbf{3} & 3 & 0.065\\ 
			\verb|^|FCHI & 5 & \textbf{4} & 1 & 0.049\\ 
			\verb|^|FTSE & 6 & \textbf{4} & 2 & 0.053\\ 
			\hline
		\end{tabular}
		\label{tab:ex:eta:Lasso}
		\caption{Example \ref{ex:eta}: Variable name, id $j$, number of occurrence as regressors, ordering function $\eta_L$ based on maximum Lasso Ordering and $5$-fold cross-validated $\lambda_j$}
	\end{table}
	If two or more variables have the same number of occurrences as regressors for other variables, we choose randomly to determine a unique ordering. If one or more variables do not occur as regressors at all, we assign them the last ranks and break ties by choosing randomly. 
\end{example}
\subsection{Sparse R-vine structure selection with the Lasso}\label{subsec:treeselection}
Knowing the ordering function $\eta$, we can write a SEM as in \eqref{eq:sem}. Assume for notational convenience that the ordering $1,\dots,d$ already reflects the ordering $\eta$ as chosen in Section \ref{subsec:calculationeta}, i.\,e.\ $\eta\left(j\right) \equiv j$. The first two equations of the SEM are trivially described. However, we can not directly use the Lasso to solve the $d-2$ later SEM equations stepwise or simultaneously. If we do, we might end up with non zero coefficients, which cannot be translated into a valid R-vine matrix as in Example \ref{ex:vine1sem}. It is much more likely that we obtain a sparse SEM as in Example \ref{ex:semworvine}, which does not have a representation as R-vine in the sense of Definition \ref{def:rvinerepsem} because of the restrictions imposed by the proximity condition. Additionally, we have to keep in mind that the solution to our SEM is also dependent on the choice of the penalization parameter $\lambda$. Thus, for different values of $\lambda$, different R-vine representations with different levels of sparsity result. We will now present an approach which computes an R-vine structure matrix $\bm{M}$ together with a coefficient matrix $\bm{\Gamma}_\lambda$, flexibly parametrizing the non-independence copulas in the R-vine in terms of $\lambda$. We consider the first R-vine tree and all higher order trees separately.

\subsubsection{Selection of the first R-vine tree $T_1$}
Let $\bm{M}$ be a $d \times d$ matrix with $\diag\left(\bm{M}\right) = \left(m_{1,1},\dots,m_{d,d}\right)$. To obtain a valid R-vine matrix, we trivially set the entry $m_{d,d-1} = m_{d,d}$ and we are left to determine Lasso regularization paths for the remaining $d-2$ columns of $\bm{M}$. Thus, we have the regression problems for $j=3,\dots,d$:
\begin{equation}\label{eq:Lassoproblemfirsttree}
	\min_{\bm{\varphi} \in \mathbb{R}^{j-1}}~\Bigg(\frac{1}{2n}\sum_{i=1}^n~\bigg(x_{i,j} - \sum_{\ell=1}^{j-1}~\varphi_{j,\ell} x_{i,\ell}\bigg)^2 + \lambda_j \sum_{\ell=1}^{j-1} \left|\varphi_{j,\ell}\right|\Bigg),
\end{equation}
and denote the solutions as $\widehat{\bm{\varphi}}_{j}^\lambda = \left(\widehat{\varphi}_{j,1}^\lambda,\dots,\widehat{\varphi}_{j,j-1}^\lambda\right) \in \mathbb{R}^{j-1}$. To formalize how we process these solutions, recall the definition of the regularization path by the set $\Lambda$ returning the non-zero coefficients in the regression of $X_j$ for each value of $\lambda \geq 0$:
\begin{equation*}
	\Lambda\left(\lambda,j\right) = \left\{\ell: \widehat{\varphi}_{j,\ell}^\lambda \neq 0 \mbox{ in } \widehat{\bm{\varphi}}_{j}^\lambda\right\},\ \mbox{with } k\left(\lambda,j\right) = \left|\Lambda\left(\lambda, j\right)\right|.
\end{equation*}
Clearly, $k\left(\lambda_1,j\right) \geq k\left(\lambda_2,j\right)$ for $\lambda_1 \leq \lambda_2$. If $\varphi_{1} \in \Lambda\left(\lambda_1,j\right)$ and $\varphi_{2} \in \Lambda\left(\lambda_1,j\right)$ but $\varphi_{1} \in \Lambda\left(\lambda_2,j\right)$ and $\varphi_{2} \notin \Lambda\left(\lambda_2,j\right)$ for $\lambda_1 < \lambda_2$, we say $\varphi_{1} \succ \varphi_{2}$. This terminology is necessary to obtain an ordering on the set $\Lambda\left(\lambda,j\right)$. It is motivated by the fact that we want to obtain the coefficients which are non-zero for the largest penalization values of $\lambda$. Thus, assume we have two coefficients for the problem \eqref{eq:Lassoproblemfirsttree}, $\widehat{\varphi}_1 = \widehat{\varphi}_2 = 0$ for some $\lambda > 0$. Now, letting $\lambda \to 0$, both coefficients will become non-zero in the end, as the penalization shrinks to zero. However, if there exists a $\lambda' > 0$ such that \eqref{eq:Lassoproblemfirsttree} is solved with $\lambda_j = \lambda'$ and we obtain $\widehat{\varphi}_1 \neq 0$ but $\widehat{\varphi}_2 = 0$, we consider $\varphi_1$ the more important coefficient and denote $\varphi_1 \succ \varphi_2$. The set $\Lambda\left(\lambda,j\right)$ contains all non-zero regressors for the penalization value $\lambda$ of the regression problem \eqref{eq:Lassoproblemfirsttree}, ordered according to their first non-zero occurrence, i.\,e.\ the regularization path. In the case of two or more $\widehat{\varphi}_j$ which are simultaneously non-zero on the regularization path, we take the one with the highest absolute value of the coefficient once they occur. This means, $\Lambda\left(\lambda,j\right)_k$ is the $k$-th non-zero regressor on the regularization path of the regression problem \eqref{eq:Lassoproblemfirsttree}. For the first R-vine tree $T_1$, let $\Lambda\left(\lambda,j\right)_k$ be the $k$-th entry in $\Lambda\left(\lambda,j\right)$ according to the ordering $\succ$. Then, $T_1$ is chosen such that
\begin{equation}\label{eq:firsttreecondition}
	\left(\kappa_1\left(3\right),\dots,\kappa_1\left(d\right)\right) = \left(\Lambda\left(0,3\right)_1,\dots,\Lambda\left(0,d\right)_1\right).
\end{equation}
Setting $\lambda = 0$ means we obtain the entire regularization path for each $j=3,\dots,d$ stored in $\Lambda\left(0,j\right)$. Together with the trivially set pair $\kappa_1\left(2\right)=1$, each pair $\left(i, \kappa_1\left(i\right)\right)$, $i=2,\dots,d$ corresponds to an edge in $T_1$. These are $d-1$ pairs and no pair can occur more than once since each left hand side of the equations is different. Thus, we set $T_1 = \left(V = \left\{1,\dots,d,\right\}, E_1 = \left\{j,\kappa_1\left(j\right)\right\}_{j=2,\dots,d}\right)$. The R-vine matrix $\bm{M}$ has the following form.
\begin{equation*}
	\bm{M} = \left(
	\begin{array}{cccccc}
		d&&&&&\\
		&d-1&&&&\\
		&&\ddots&&&\\
		&&&3&&\\
		&&&&2&\\
		\kappa_1\left(d\right)&\kappa_1\left(d-1\right)&\dots&\kappa_1\left(3\right)&1&1
	\end{array}
	\right)
\end{equation*}
Thus, also the sets $\RRR_1\left(\eta\left(j\right)\right) = \left\{\kappa_1\left(j\right)\right\}$ for $j=2,\dots,d$ are updated. As mentioned, in this step we calculate the entire regularization path for each $X_j$, $j=3,\dots,d$ with respect to \eqref{eq:Lassoproblemfirsttree}. However, we can not be sure if in one path subsequent values adhere to the proximity condition, see Example \ref{ex:semworvine}. We keep the paths stored as they may be compatible with the proximity condition which we will check later on and which may save computation time. Recall that the regularization paths also include the corresponding $\lambda$ for which the coefficients on the regularization path become non zero. This finishes the selection of $T_1$.

\subsubsection{Selection of the higher order trees $T_2,\dots,T_{d-1}$}
In the first tree, it was not necessary to take into account the proximity condition to compute a valid R-vine matrix $M$. However, for the sequential steps, this will be the case. We consider again the $6$-dimensional data from Example \ref{ex:eta}.
\begin{example}[Example \ref{ex:eta} cont.]\label{ex:blackwhitelist}
	We use the ordering function $\eta$ to obtain $\left(\eta\left(1\right),\dots,\eta\left(6\right)\right) = \left(6,5,4,1,2,3\right)$. We set the value $m_{6,5} = m_{6,6}$ as it is the only allowed entry. Computing the regularization paths for the variables $\left(\eta\left(3\right),\dots,\eta\left(6\right)\right) =  \left(4,1,2,3\right)$, i.\,e.\ solutions to \eqref{eq:Lassoproblemfirsttree}, we obtain:
	\begin{equation*}
		\begin{aligned}
			\Lambda\left(0, 4\right) &= \left\{5,6\right\} & \Rightarrow \kappa_1\left(4\right) = 5,\\
			\Lambda\left(0, 1\right) &= \left\{4,5,6\right\} & \Rightarrow \kappa_1\left(1\right) = 4,\\
			\Lambda\left(0, 2\right) &= \left\{5,4,1,6\right\} & \Rightarrow \kappa_1\left(2\right) = 5,\\
			\Lambda\left(0, 3\right) &= \left\{2,6,1,4,5\right\} & \Rightarrow \kappa_1\left(3\right) = 2.
		\end{aligned}
	\end{equation*}
	Note here that we consider $\lambda_j = 0,\ j = 4,1,2,3$ as we want to obtain the entire path without any shrinkage. We take the first coefficients according to the ordering $\succ$ to determine the first R-vine tree $T_1$, encoded by the $d$-th row of the partial R-vine matrix $\bm{M}'$. 
	\begin{tabular}{ll}
		\begin{minipage}[l]{0.45\textwidth}
			\begin{equation*}
				\bm{M}' = \left(
				\begin{array}{cccccc}
					3&&&&&\\
					&2&&&&\\
					&&1&&&\\
					&&&4&&\\
					\Box&&&6&6&\\
					2&5&4&5&5&5
				\end{array}
				\right)
			\end{equation*}
		\end{minipage}
		&
		\hspace{-0.5cm}
		\begin{minipage}[l]{0.45\textwidth}
			\begin{equation*}
				\bm{M} = \left(
				\begin{array}{cccccc}
					3&&&&&\\
					6&2&&&&\\
					1&6&1&&&\\
					4&1&6&4&&\\
					5&4&5&6&6&\\
					2&5&4&5&5&5
				\end{array}
				\right)
			\end{equation*}
		\end{minipage}
	\end{tabular}\\[0.25cm]
	We need to determine the second tree, i.\,e.\ $\bm{M}'_{d-1,}$. First, we note that $m_{5,4}=6$ is the only valid choice. For the general case, consider the missing entry $m_{5,1}$, marked by $\Box$. First, we check whether the second entry in the regularization path, $\Lambda\left(0,3\right)_2 = 6$ is valid. By checking the proximity condition \eqref{eq:proximitycondition3}, this is not the case as $2$ and $6$ are not connected in $T_1$. Thus, we recompute the regularization path such that $\Lambda\left(0,3\right)_1 = m_{6,1} = 2$ and $\Lambda\left(0,3\right)_2$ adheres to the proximity condition. The set of possible regressors are the entries on the main diagonal to the right of the first column $2,1,4,6,5$, where $2$ is already occurring. This leaves us with $1,4,6,5$. From these, only $5$ is a possible entry according to the proximity condition. Thus, the remaining $1,4,6$ are set on a \textit{blacklist set} for the entry $m_{5,1}$ by $\BBB\left(5,1\right) = \left\{1,4,6\right\}$. Next, we re-run the penalized regression to find a new regularization path reflecting the blacklist. However, we also have to include that there are regressors we want to include on the regularization path before the second regressor, i.\,e.\ $m_{6,1} = 2$. We will call it the \textit{whitelist set} $\WWW\left(5,1\right)=\left\{m_{6,1}\right\} =\left\{2\right\}$. Since we can set individual penalties for each variable, we set $\lambda_{3,2} = 0$. The optimization problem for the entry $m_{5,1}$ is given by:
	\begin{equation*}
		\min_{\bm{\varphi} \in \mathbb{R}^{1}}~\Bigg(\frac{1}{2n}\sum_{i=1}^n~\bigg(X_{i,3} - \sum_{\ell \in \left\{1,2,4,5,6\right\} \setminus \left\{1,4,6\right\}}~\varphi_{3,\ell} X_{i\ell}\bigg)^2 \Biggr.  
		+ \sum_{\ell \in \left\{\left\{1,2,4,5,6\right\} \setminus \left\{1,4,6\right\}\right\} \setminus 2} \lambda_{3,\ell}\left|\varphi_{3,\ell}\right|\Bigg).
	\end{equation*}
	Thus, we obtain a new sequence $\Lambda\left(0,3\right)$ such that $\Lambda\left(0,3\right)_1=m_{6,1}$ and $\Lambda\left(0,3\right)_2=5$ adheres to the proximity condition. Whenever we have to start a new regression since the next regressor on the regularization path does not adhere to the proximity condition as described previously, we denote this as a \textit{proximity condition failure (pcf)}. In the end, we obtain the complete R-vine matrix $\bm{M}$. Additionally, we yield the corresponding $\lambda$ entries for each entry, based either on an already computed regularization path or a new computation. We store it together with the R-vine matrix.
\end{example}
Using this approach, we complete a partial R-vine matrix column-wise from right to left in $d-1$ steps. However, since each lower order tree put restrictions on higher order trees by the proximity condition, we have $j$ iterations in the $d-j$-th column for $j=1,\dots,d-1$. From a computational point of view, it is more favourable to complete the matrix row-by-row, i.\,e.\ tree by tree. Thus, the structure estimation, i.\,e.\ computation of regularization paths, can be done in parallel. Because of the particular importance, we restate the optimization leading to the higher order tree estimates in the general form.

\begin{Definition}[Higher order tree selection]\label{def:higherordertreeoptimization}
	Let $\bm{M}$ be a partial R-vine matrix and assume without loss of generality the main diagonal $\left(m_{1,1},\dots,m_{d,d}\right)  =\left(d,\dots,1\right)$. For each matrix entry $m_{i,j}$ with $i > j$, define the set of potential regressors\\ $\HHH\left(i,j\right) = \left\{m_{j+1,j+1},\dots,m_{d,d}\right\}$, the whitelist $\WWW\left(i,j\right) = \left\{m_{d,j},\dots,m_{i+1,j}\right\}$ and the blacklist\\ $\BBB\left(i,j\right) = \left\{\ell \in \HHH\left(i,j\right) \setminus \WWW\left(i,j\right): \ell \mbox{ does not satisfy the pc.} \right\}$. We solve for $\bm{\varphi} \in \mathbb{R}^{j-1 - \left|\BBB\left(i,j\right)\right|}$ the optimization problem:
	\begin{equation}\label{eq:defhigherordertree}
		\min_{\bm{\varphi}}~\Bigg(\frac{1}{2n}\sum_{k=1}^n~\bigg(X_{k,j} - \sum_{\ell \in \HHH\left(i,j\right) \setminus \BBB\left(i,j\right)}~\varphi_{j,\ell} X_{k,\ell}\bigg)^2 + \sum_{\ell \in \left(\HHH\left(i,j\right) \setminus \BBB\left(i,j\right)\right) \setminus \WWW\left(i,j\right)} \lambda_{j,\ell}\left|\varphi_{j,\ell}\right|\Bigg),
	\end{equation}
	to obtain a regularization path $\Lambda\left(\lambda,j\right)$ such that
	\begin{itemize}
		\item $\Lambda\left(0,j\right)_\ell = m_{d-\ell+1,j}$ for $\ell \in 1,\dots,\left|\WWW\left(m_{i,j}\right)\right|$,
		\item $\Lambda\left(0,j\right)_{d-i+1}$ adheres to the proximity condition.
	\end{itemize}
\end{Definition}
To check whether a specific regressors $m_{i,j}$ is in the blacklist or not, we can use the partial R-vine matrix to see if \eqref{eq:proximitycondition3} holds for this value. 
This concludes the part where we deal with the structure selection of the R-vine. 
We continue with considering the sparsity, i.\,e.\ how to use the Lasso to not only calculate a feasible structure but also perform model selection. Thus, we aim to make our R-vine model sparser by setting independence copulas.

\subsection{Calculating R-vine regularization paths}\label{sec:rvineregpaths}
From the previous calculations, we obtain an R-vine structure together with a regularization path, i.\,e.\ a functional relationship between $\lambda > 0$ and the non-zero regression coefficients.
Now, we use this information to define the entire regularization path of the regression of $X_{m_{j,j}}$ onto $X_\ell$, $\ell = m_{j+1,j},\dots,m_{d,j}$ where $\bm{M}$ denotes the R-vine structure matrix. This path will be called \textit{column regularization path}. For notational convenience, we reverse the order of the rows of the matrix to obtain a new matrix $\bm{M}^*$. By this convention, the corresponding $i$-th entry in column $j$ corresponds to the $i$-th R-vine tree and we have $m^*_{i,j} = m_{d-i+1,j}$ for $j=1,\dots,d-1, i = 1,\dots,d-j$. For example, the first column of the R-vine Matrix $\bm{M}$ from Example \ref{ex:blackwhitelist} is $\bm{M}_{,1} = \left(3,6,1,4,5,2\right)$. Thus, $\bm{M}^*_{,1} = \left(2,5,4,1,6,3\right)$. Finally note that the $j$-th column in $\bm{M}$ and $\bm{M}^*$ has exactly $d-j$ non-zero entries.

\begin{Definition}[Column regularization path]\label{def:columregpath}
	Let $\bm{M}$ be an R-vine structure matrix in $d$ dimensions. A column regularization path of the reversed $j$-th column $\bm{M}^*_{,j} = \left(m^*_{1,j},\dots,m^*_{d-j,j}\right)$ is a vector $\bm{\lambda}_j = \left(\lambda_{1,j},\dots,\lambda_{d-j,j}\right) \geq 0$ for $j=1,\dots,d-2$ such that
	\begin{equation*}
		\left\{\ell: \lambda_{\ell,j} < \lambda'\right\} = \left\{\ell: c_{j,\kappa_{\ell}\left(j\right)|\kappa_{1}\left(j\right),\dots,\kappa_{\ell-1}\left(j\right)} = 0 \mbox{ in R-vine tree } T_\ell\right\}
	\end{equation*}
	for some $\lambda' > 0$.
\end{Definition}
Thus, each column $j=1,\dots,d-1$ of the R-vine matrix is assigned a vector $\bm{\lambda}_j \in \mathbb{R}^{d-j}$ which contains \textit{threshold} values. These values are a by-product of the penalized regressions we ran and specify for which threshold of penalization, the corresponding SEM coefficients are set to zero, and hence, pair copulas are set to independence copulas. Thus, only by comparing component-wise $\bm{\lambda}_j > \lambda'$ for some $\lambda' > 0$, the column regularization path helps to set pair copulas to the independence copula to reflect a specific degree of sparsity associated to $\lambda'$. For the column $d-1$ where we only have one value, we perform a single regression, so called \textit{soft thresholding} to calculate the corresponding value of $\lambda_{d-1} \geq 0$.\\
The advantage of this path is now that we are able to regularize each column of the R-vine matrix independently based on a solid theoretical reasoning, i.\,e.\ the Lasso. In practice, we consider the R-vine family matrix $\Gamma$ and fix a specific threshold of $\lambda' > 0$. 
We consider the column regularization path $\bm{\lambda}_j$ and calculate component-wise the $j$-th column of the R-vine family matrix $\Gamma$ as
\begin{equation*}
	\left(\Gamma_{d,j},\dots,\Gamma_{d-j,j}\right) = \left(\mathds{1}_{\left\{\lambda_{1,j} \geq \lambda'\right\}},\dots,\mathds{1}_{\left\{\lambda_{d-j,j} \geq \lambda'\right\}}\right)
\end{equation*}
Note that we reverse the ordering to work solely with lower triangular matrices, i.\,e.\ $\Gamma_{d-i+1,j}$ corresponds to $\lambda_{i,j}$ for $i = 1,\dots,d-j$. Thus, all coefficients which are on the regularization path associated to a value of $\lambda < \lambda'$, are set to zero, and hence, the corresponding pair copula is set to the independence copula. The remaining pair copulas are then subject to further estimation. We can not only calculate single column regularization paths, but the entire \textit{regularization path of the R-vine}.

\begin{Definition}[Regularization path of an R-vine]\label{def:rvineregpath}
	Let $\bm{M}$ be an R-vine structure matrix in $d$ dimensions. The regularization path of the R-vine is a matrix $\bm{\varLambda} \in \mathbb{R}^{d \times d}$ such that its columns $\bm{\varLambda}_{,j}$ are column regularization paths of the corresponding R-vine matrix columns $\bm{M}_{,j}$, $j=1,\dots,d-1$.
\end{Definition}

Summarizing, we obtain an R-vine structure which is not only entirely independent of pseudo-observations of lower level trees as compared to Dissmann's algorithm. It is also independent of a specific penalization level $\lambda$ since it is built stepwise by considering the non-zero regressors along the entire regularization paths for the R-vine tree sequence. This allows us to calculate one specific R-vine structure and then consider it under arbitrary many penalization levels $\lambda$, obtaining different levels of sparsity for one generally valid R-vine structure. We consider the regularization path matrix of the $6$ dimensional example R-vine from Example \ref{ex:regularizationpath} before proposing methods for choosing $\lambda$.

\begin{example}[Example \ref{ex:blackwhitelist} cont.]\label{ex:regularizationpath}
	Consider the R-vine matrix $\bm{M}$ from Example \ref{ex:blackwhitelist}. For comparison, we also display the R-vine matrix generated from Dissmann's algorithm, calculated with \texttt{VineCopula} R-package, see \cite{VineCopula}.\\
	\begin{tabular}{ll}
		\begin{minipage}[l]{0.45\textwidth}
			\begin{equation*}
				\bm{M}_{Lasso} = \left(
				\begin{array}{cccccc}
					3&&&&&\\
					6&2&&&&\\
					1&6&1&&&\\
					4&1&6&4&&\\
					5&4&5&6&6&\\
					2&5&4&5&5&5
				\end{array}
				\right)
			\end{equation*}
		\end{minipage}
		&
		\hspace{-0.5cm}
		\begin{minipage}[l]{0.45\textwidth}
			\begin{equation*}
				\bm{M}_{Dissm.} = \left(
				\begin{array}{cccccc}
					3&&&&&\\
					6&2&&&&\\
					1&6&1&&&\\
					4&1&6&4&&\\
					5&4&5&6&5&\\
					2&5&4&5&6&6
				\end{array}
				\right)
			\end{equation*}
		\end{minipage}
	\end{tabular}\\[0.25cm]
	We compute the following regularization path matrix $\bm{\varLambda}$ for the R-vine structure $\bm{M}$.
	\begin{equation*}
		\bm{\varLambda} = \left(
		\begin{array}{cccccc}
			0.0000 & 0.0000 & 0.0000 & 0.0000 & 0.0000 & 0.0000 \\ 
			0.0072 & 0.0000 & 0.0000 & 0.0000 & 0.0000 & 0.0000 \\ 
			0.0082 & 0.0039 & 0.0000 & 0.0000 & 0.0000 & 0.0000 \\ 
			0.0005 & 0.0091 & 0.4993 & 0.0000 & 0.0000 & 0.0000 \\ 
			0.0538 & 0.0210 & 0.6601 & 0.1344 & 0.0000 & 0.0000 \\ 
			0.3171 & 0.3117 & 0.7244 & 0.9481 & 0.9378 & 0.0000 \
		\end{array}
		\right)
	\end{equation*}
	We observe that the values are column-wise monotonically decreasing if there is no proximity condition failure (pcf). For example, in the first column, the original regularization path $\Lambda\left(0,3\right) = \left(2,6,1,4,5\right)$ did not meet the proximity condition and was recalculated. Thus, the values of $\lambda$ are not necessarily decreasing. We visualize the column regularization path of column $2$ with a step function, indicating the corresponding entries in the R-vine matrix.
	\begin{figure}[h]
		\centering
		\includegraphics[width=0.5\textwidth]{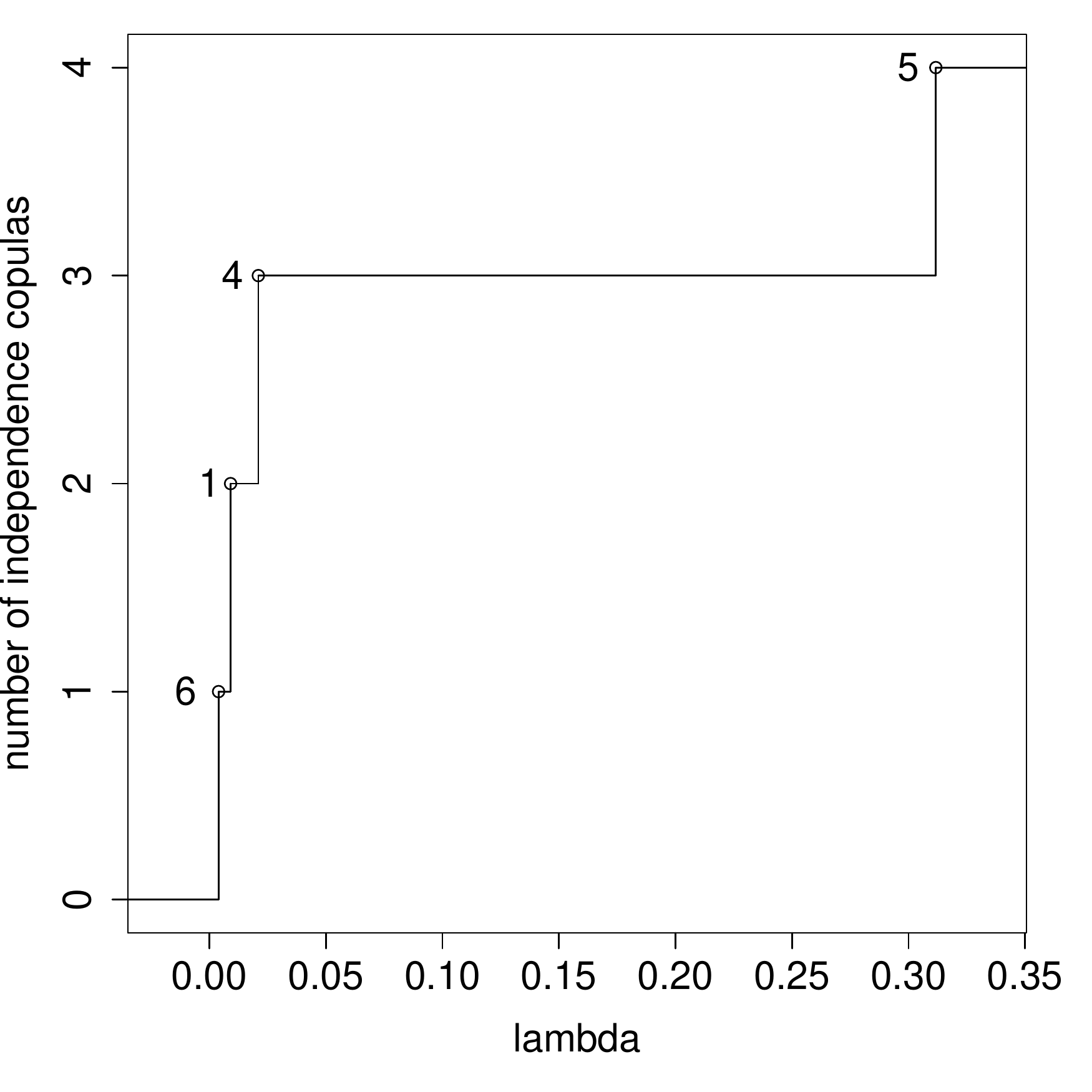}
		\caption{Column regularization path of column $2$}
		\label{fig:ex:stepfun}
	\end{figure}
	As $\lambda \to 0$, more and more pair copulas are set to the independence copula, starting from higher order trees to lower order tres. The matrix $\bm{\varLambda}$ can now be used to regulate the sparsity, i.\,e.\ the number of independence copulas in our R-vine model.
\end{example}

\subsection{Selection of the tuning parameter $\lambda$}\label{subsec:selectlambda}
We propose two approaches how we can utilize the regularization path matrices to obtain sparse R-vine models. A high value in these matrices means a significant contribution to the model fit, where a low values means the opposite. Introducing now a threshold value $\lambda_T$ and checking whether or not entries in $\bm{\varLambda}$ are below or above this value, the corresponding entries in $\Gamma$ are set to the independence copula or left for estimation of the pair copula type and parameter. Denote the regularization path of the R-vine by $\bm{\varLambda} = \left(\lambda_{i,j}\right)_{i=1,\dots,d; j=1,\dots,d}$ with family matrix $\bm{\Gamma} = \left(\gamma_{i,j}\right)_{i=1,\dots,d; j=1,\dots,d}$.

\subsubsection{Single threshold approach}
The first approach is to specify some threshold $\lambda_T > 0$ and calculate the family matrix entries according to
\begin{equation}\label{eq:thresholding1}
	\gamma_{i,j} = \mathds{1}_{\left\{\lambda_{i,j} \geq \lambda_T\right\}},\ j=1,\dots,d-1,\ i = 1,\dots,d-j.
\end{equation}
Pair copulas corresponding to unit entries in the family matrix are then subject to, e.\,g.\ maximum likelihood estimation. Such an approach can easily be evaluated using a grid of $\lambda_T \in \left(0,1\right)$. Recall that we only need to compute the structure and regularization matrix once upfront and then evaluate the corresponding threshold. In the data example, we consider a grid of threshold parameters.


\subsubsection{Adaptive threshold approach}
A second approach is to specify not a threshold value itself, but to calculate the threshold such that a specified share of the entries in $\bm{\varLambda}$ fall below the threshold. Recall that $\bm{\varLambda}$ has $\binom{d}{2}$ entries as lower diagonal matrix. Our intention is to grab the highest $100\mu \%$ of the values in $\bm{\varLambda}$. Thus, we solve the following equation for a threshold $\lambda_\mu$:
\begin{equation}\label{eq:thresholding3}
	\lfloor\mu \cdot \binom{d}{2}\rfloor = \sum_{j=1,\dots,d-1, i = 1,\dots,d-j}^d~\mathds{1}_{\left\{\lambda_{i,j} \geq \lambda_\mu\right\}}
\end{equation}
This threshold can easily be found by sorting all entries of $\bm{\varLambda}$ decreasingly and stop once $\lfloor\mu \cdot \binom{d}{2}\rfloor$ entries have been found.

\begin{example}[Example \ref{ex:regularizationpath} cont.]\label{ex:lambdaselection}
	We consider the regularization path matrix $\varLambda$ as in Example \ref{ex:regularizationpath}. For the single threshold approach, we choose $\lambda_T = 0.1$ to obtain $\bm{\Gamma}_1$ and for the adaptive threshold approach, we use $\mu = 0.5$.
	With $\binom{d}{2} = \binom{6}{2} = 15$, we have $\lfloor\mu \cdot \binom{6}{2}\rfloor = 7$, i.\,e.\ we select the entries with the highest $7$ values in $\bm{\varLambda}$, obtaining $\bm{\Gamma}_2$.\\
	\begin{tabular}{cc}
		\begin{minipage}[l]{0.5\textwidth}
			\begin{equation*}
				\bm{\Gamma}_1 = 
				\left(
				\begin{array}{cccccc}
					-&&&&&\\
					0&-&&&&\\
					0&0&-&&&\\
					0&0&1&-&&\\
					0&0&1&1&-&\\
					1&1&1&1&1&-
				\end{array}
				\right)
			\end{equation*}
		\end{minipage}
		&
		\hspace{-0.0cm}
		\begin{minipage}[l]{0.5\textwidth}
			\begin{equation*}
				\bm{\Gamma}_2 = 
				\left(
				\begin{array}{cccccc}
					-&&&&&\\
					0&-&&&&\\
					0&0&-&&&\\
					0&0&1&-&&\\
					0&0&1&0&-&\\
					1&1&1&1&1&-
				\end{array}
				\right)
			\end{equation*}
		\end{minipage}
	\end{tabular}
\end{example}

\section{Numerical examples}\label{sec:numericalexamples}
\subsection{Simulation study}\label{subsec:simstudy}
We demonstrate the overall feasibility of our proposed approach, and superiority compared to the current standard algorithm for selection of R-vines. We show that the Lasso outperforms Dissmann's method in terms of the modified BIC, see \eqref{eq:mbic}, when the data is sparse. Additionally, our approach is much faster and allows to separate the structure selection from the actual pair copula estimation. Thus, one structure matrix together with its R-vine regularization path matrix can be used to infer arbitrarily many different sparse R-vine models.\\
We gathered data from the S\&P100 constituents from January 01, 2013 to December 31, 2016. Removing incomplete data because stocks entering or leaving the index, we obtain $d = 85$ dimensions on $n = 1007$ observations. We calculated daily log-returns of the adjusted closing prices, incorporating dividends and stock splits. We fitted \textit{ARMA-GARCH}$\left(p,q\right)$ models with $p,q \in \left\{0,1\right\} \times \left\{0,1\right\}$ and residuals distributed according to a normal, Student-t or skew Student-t distribution onto each of the $85$ time series, obtaining $12$ candidate models for each time series. We chose the time series model with highest log-Likelihood and calculated the corresponding standardized residuals. These residuals are transformed to the \textit{u-scale} using their empirical cumulative distribution function. We use Dissmann's algorithm to fit different models with several degrees of sparsity by imposing $2, 5, 10$-truncations in the model fit. All pair copula families implemented in the R-package \texttt{VineCopula} were allowed and a level $\alpha = 0.05$ independence test was performed. Thus, we obtain three scenarios $S_1,S_2,S_3$ from which we draw $M = 50$ replications of $n = 1000$ samples each. In all these scenarios, there is clearly non-Gaussian dependence, as we have e.\,g.\ $89$ Student $t$-copulas and $29$ Frank copulas out of $167$ non independence copulas total. The proportions are very similar in the $5$-- and $10$-- truncated scenarios. For these replicated data sets, we fit Dissmann's approach using the \texttt{VineCopula} R-package. Additionally, we use our novel Lasso approach and use the single threshold approach with $\lambda_T = \left(1/4\right)^4$, $\lambda_T = \left(1/5\right)^4$ and adaptive threshold approach with $\mu = 0.1$ and $\mu = 0.2$. We additionally test for independence copulas using a significance level $\alpha = 0.05$ after applying the threshold.
To draw conclusions, we consider boxplots where we compare the true values with both our Lasso approaches and Dissmann's algorithm. 
\begin{figure}[ht]
	\centering
	\includegraphics[width=0.32\textwidth, trim={0.1cm 0.1cm 0.1cm 0.1cm},clip]{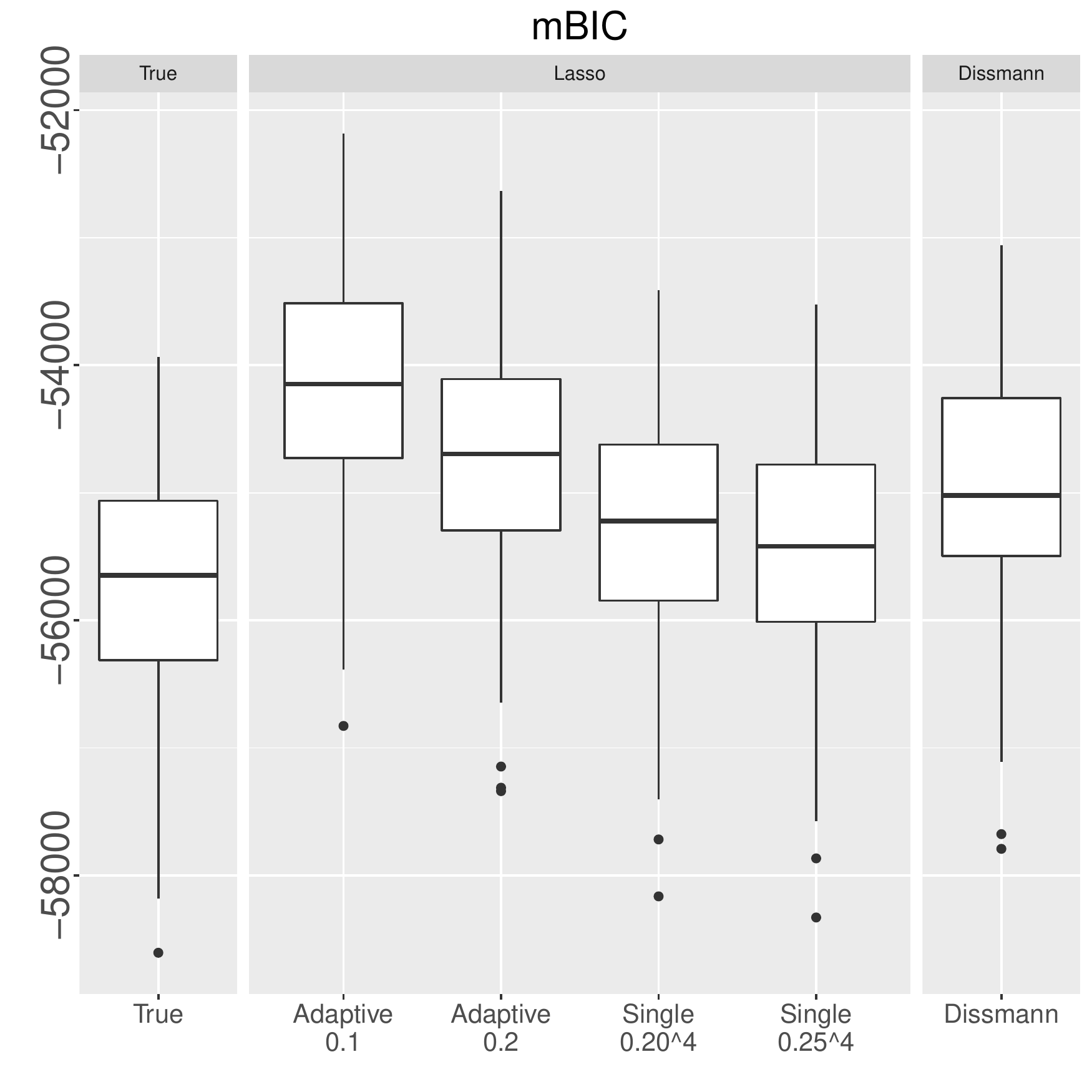}
	\includegraphics[width=0.32\textwidth, trim={0.1cm 0.1cm 0.1cm 0.1cm},clip]{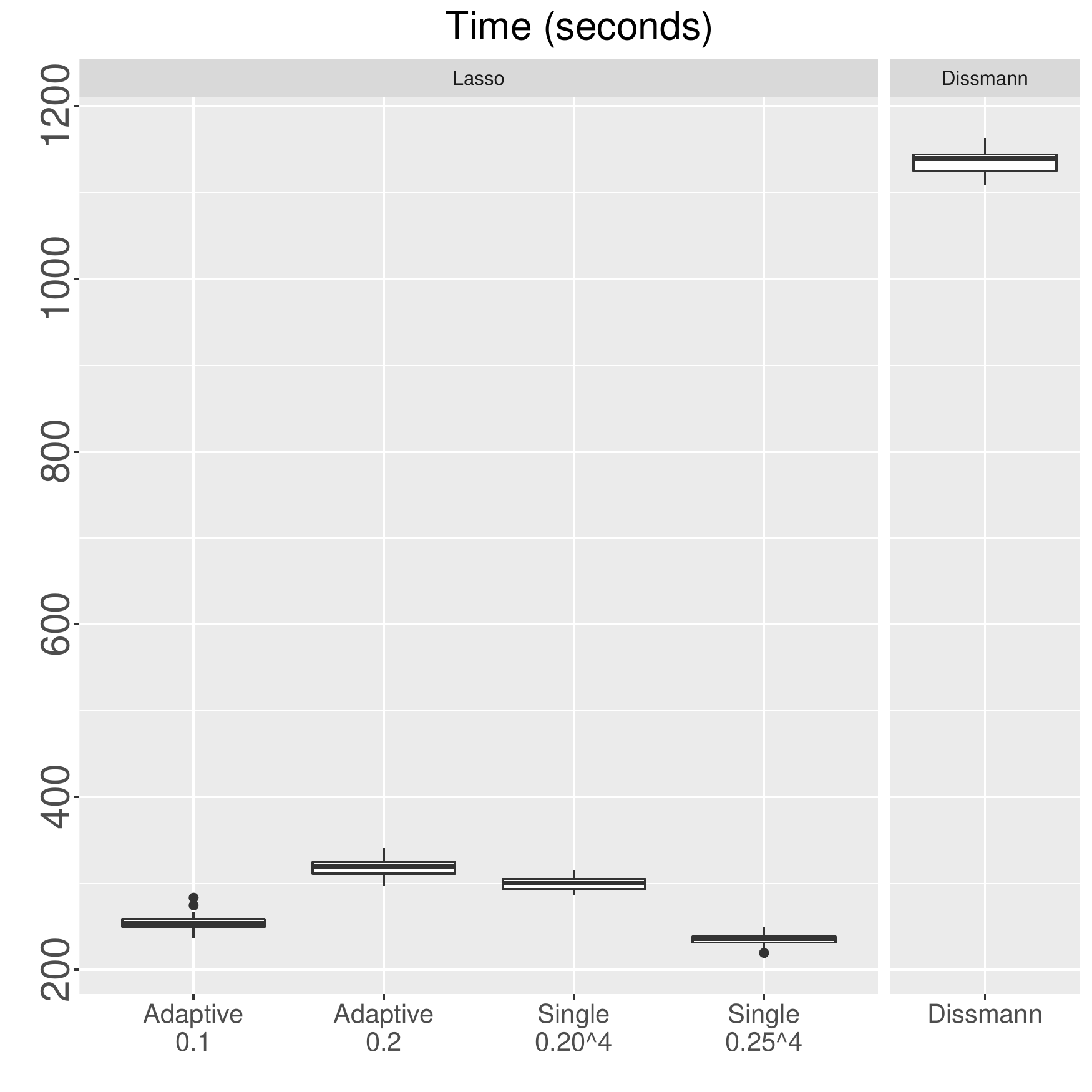}
	\includegraphics[width=0.32\textwidth, trim={0.1cm 0.1cm 0.1cm 0.1cm},clip]{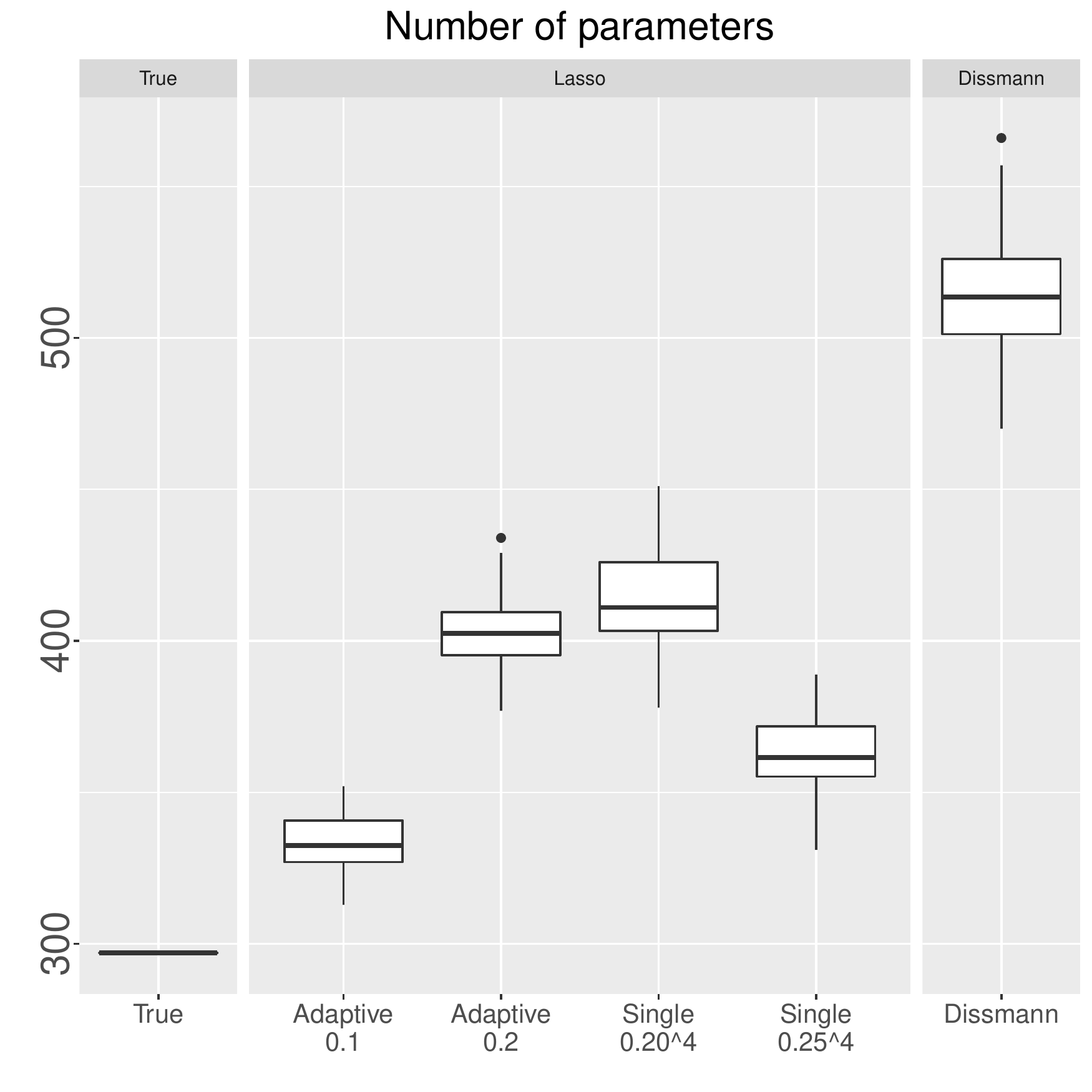}
	\caption{Scenario $S_1$: Comparison of Lasso approach for fixed and flexible thresholding with true model and Dissmann's algorithm considering mBIC, computation time in seconds and number of parameters on $50$ replications (from left to right)}
	\label{fig:simstudy:results1}
\end{figure}
The boxplots in Figure \ref{fig:simstudy:results1} show mBIC, computation time and number of parameters for scenario $S_1$. The remaining plots for the other scenarios are similar and hence deferred to Appendix \ref{sec:app:simstudy}.
We see that our approach attains mBIC closer to the true model than Dissmann in all the scenarios. Our novel approaches require much less parameter, where the single threshold approach is superior to the adaptive threshold approach. In terms of computation time, the single threshold approach is also advantageous to its competitors. This is particularly surprising since the single threshold is the same for all scenarios and works for different degrees of sparsity. We stress again that once a Lasso structure and regularization path is found, multiple models can be considered by varying the thresholding parameter $\lambda_T > 0$. 

\subsection{Data example}\label{subsec:dataapplication}
\begin{figure}[ht]
	\centering
	\includegraphics[width=0.45\textwidth, trim={0.1cm 0.3cm 0.7cm 0.3cm},clip]{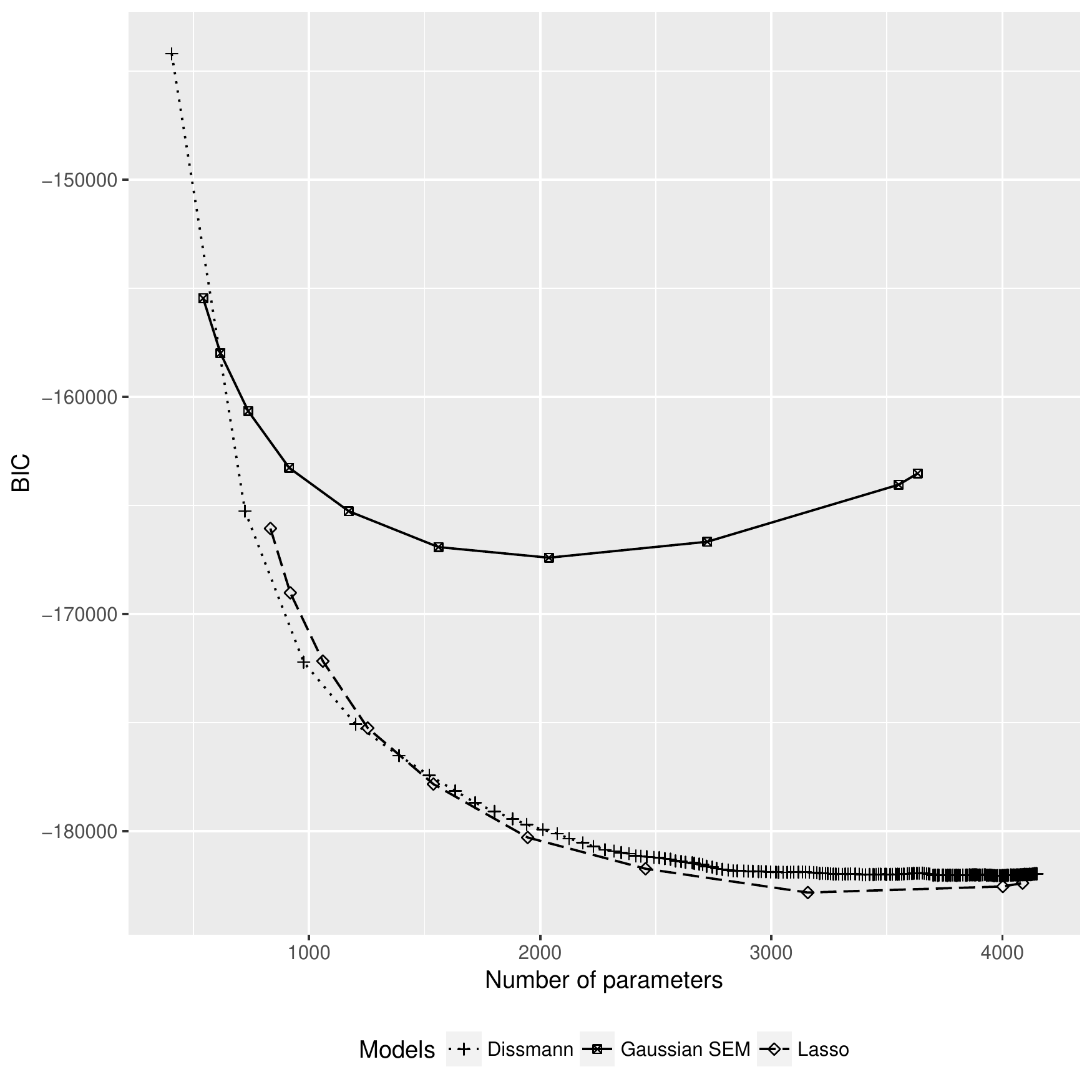}
	\includegraphics[width=0.45\textwidth, trim={0.1cm 0.3cm 0.7cm 0.3cm},clip]{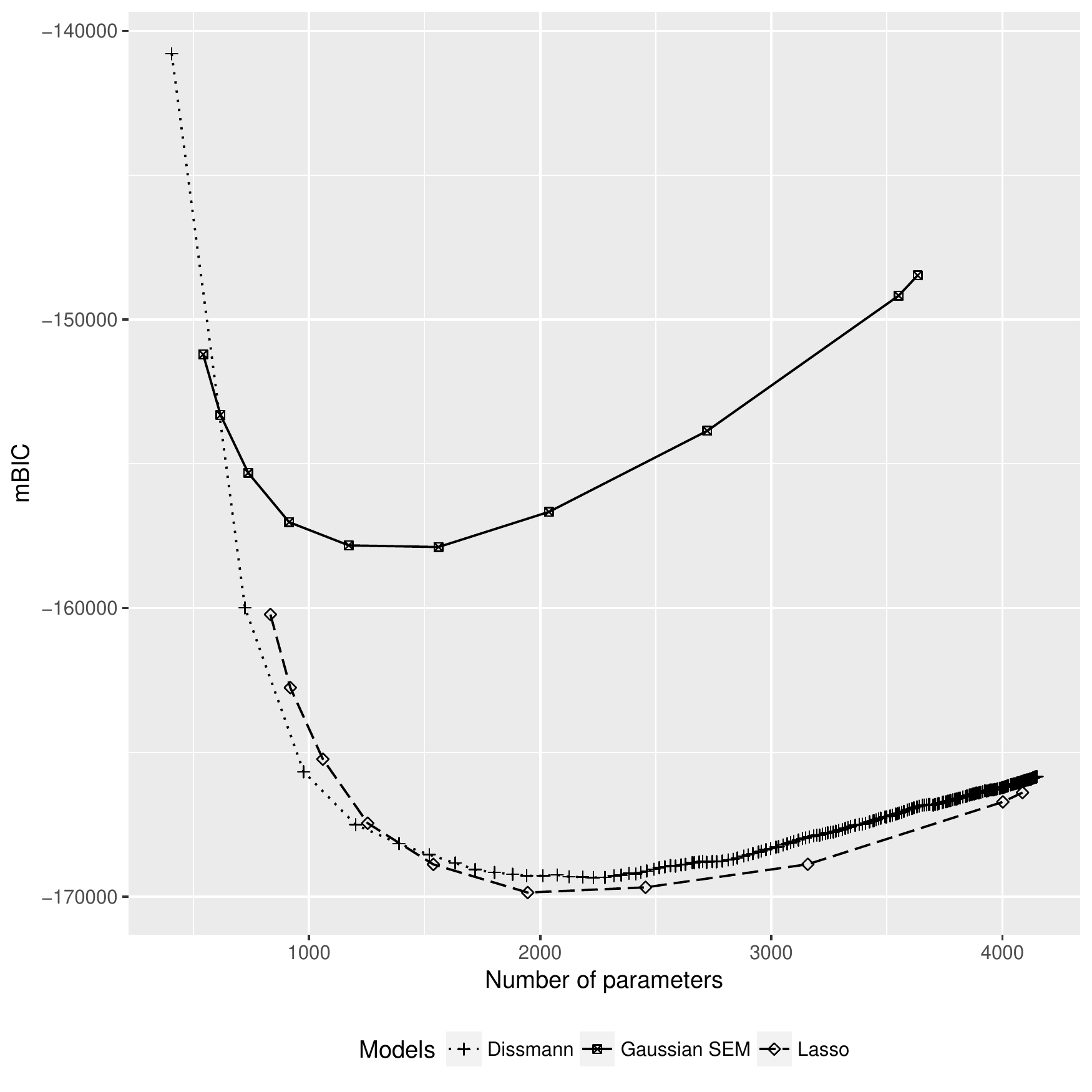}
	\caption{Comparison of Lasso approach with single threshold $\lambda_T \in \left\{0.05^4, 0.1^4,\dots, 0.45^4, 0.5^4\right\}$ vs. Gaussian SEM model with same threshold vs. $t$-truncated Dissmann's algorithm, $t=1,\dots,221$ on \textit{u-scale} by number of parameters vs. BIC (left) and number of parameters vs. mBIC (right)}
	\label{fig:application}
\end{figure}

We scale our approach to even higher dimensions. Because of the availability of data, we again consider a financial dataset. Thus, we obtain data from the S\&P500 constituents, also from January 1, 2013 to December 31, 2016. We isolate $d = 222$ stocks which fall into the sectors Financial Services $(70)$, Health Care $(40)$, Industrials $(54)$, Information Technology $(52)$ and Telecommunication Services $(6)$. We apply the same procedure as to our data prepared for the simulation study and use suitable \textit{ARMA-GARCH} models to remove trends and seasonality from the time series. The residuals are then transformed using the empirical cumulative distribution function to the \textit{u-scale}. To obtain models, we use Dissmann's algorithm with a level $\alpha = 0.05$ independence test and $1,\dots,221$-truncation, i.\,e.\ we fit a full model and the split it into the first $k$ trees for $k=1,\dots,221$ to obtain submodels. We consider only one-parametric pair copula families and the $t$-copula. The same pair copula selection also applies for our approach where we calculated models along a grid of single threshold values $\lambda_T \in \left\{0.05^4, 0.1^4,\dots, 0.45^4, 0.5^4\right\}$. We additionally also test for independence using a significance level $\alpha = 0.05$. As a comparison, we also include the merely Gaussian SEM. We plot the corresponding BIC and mBIC values of both models, see Figure \ref{fig:application}.
For the Lasso approach, the BIC and mBIC of the models is decreasing with decreasing $\lambda_T$ as less pair copulas get penalized and we obtain more and more parameters. We see that BIC and mBIC attain a minimum for both Lasso and Dissmann approach different from the full models, where $\lambda_T^{opt} = (1/4)^4$ and $t^{opt} = 16$ for Dissmann. Both our Lasso approach and Dissmann's algorithm outperform the Gaussian SEM significantly which indicates non-Gaussian dependence. Additionally, we see that the Lasso outperforms the Dissmann method in both criteria as it attains smaller values. For computation times, we report that one fit of the Lasso approach took approximately 30 minutes. The entire Dissmann fit for the full model took over 5,5 hours, all times on a Linux Cluster with 32 cores. Thus, all of the Lasso models on the grid were fitted before the full Dissmann fit was complete. The Gaussian-SEM is much faster compared to both non-Gaussian approaches. We see that in terms of mBIC, the optimal models have around $2,000$ parameters out of total $222\times \left(221\right)/ 2 \approx 25,000$ parameters. We observe most often Student $t$ (386) and Frank copulas (810). Our expectation is that for higher dimensional models, the ratio of significant parameters to total number of parameters becomes even smaller, making sparse model selection key for high dimensional setups.

\section{Discussion}\label{sec:discussion}
We presented an entirely novel structure selection method for high dimensional vine copulas. Our proposal is based on application of the well known Lasso in the context of dependence modeling. We described the theoretical connection via structural equation models and how we can adapt the Lasso to reflect the proximity condition, a key ingredient for vine models. We transferred the concept of regularization paths to vine copulas and proposed methods for finding thresholds for models. In our numerical examples, we demonstrated the feasibility and superiority of our approach over existing methods when non-Gaussian dependence is present. We observed superior fit with respect to stronger penalizing goodness-of-fit measures and for computation time. We believe that in especially high dimensional settings, it is of paramount interest to first rule out the majority of the unnecessary information to only obtain the most significant contributions, which is clearly a characterizing feature of the Lasso. However, this also depends on the choice of the tuning parameter $\lambda$. More elaborate selection strategies for $\lambda$ and other penalty functions, e.\,g.\ the \textit{elastic net} by \cite{ZouHastie2005}, are also part of current and future research.

	\section*{Acknowledgement}\label{sec:acknowledgment}
	The first author is thankful for support from Allianz Deutschland AG. The second author is supported by the German Research foundation (DFG grant GZ 86/4-1). Numerical computations were performed on a Linux cluster supported by DFG grant INST 95/919-1 FUGG.

%
%
%
%

\bibliographystyle{chicago}
\bibliography{bibliography}

\begin{thebibliography}{}

\bibitem[\protect\citeauthoryear{Aas}{Aas}{2016}]{econometrics4040043}
Aas, K. (2016).
\newblock Pair-copula constructions for financial applications: A review.
\newblock {\em Econometrics\/}~{\em 4\/}(4).

\bibitem[\protect\citeauthoryear{Aas, Czado, Frigessi, and Bakken}{Aas
  et~al.}{2009}]{Aasetal2009}
Aas, K., C.~Czado, A.~Frigessi, and H.~Bakken (2009).
\newblock Pair-copula constructions of multiple dependence.
\newblock {\em Insurance, Mathematics and Economics\/}~{\em 44}, 182--198.

\bibitem[\protect\citeauthoryear{Akaike}{Akaike}{1973}]{Akaike1973}
Akaike, H. (1973).
\newblock Information theory and an extension of the maximum likelihood
  principle.
\newblock In B.~N. Petrov and F.~Csaki (Eds.), {\em Proceedings of the Second
  International Symposium on Information Theory Budapest}, Akademiai Kiado,
  pp.\  267--281.

\bibitem[\protect\citeauthoryear{Bauer and Czado}{Bauer and
  Czado}{2016}]{BauerCzado2016}
Bauer, A. and C.~Czado (2016).
\newblock Pair-{C}opula {B}ayesian networks.
\newblock {\em Journal of Computational and Graphical Statistics\/}~{\em
  25\/}(4), 1248--1271.

\bibitem[\protect\citeauthoryear{Bedford and Cooke}{Bedford and
  Cooke}{2001}]{BedfordCooke2001}
Bedford, T. and R.~Cooke (2001).
\newblock Probability density decomposition for conditionally dependent random
  variables modeled by vines.
\newblock {\em Annals of Mathematics and Artificial Intelligence\/}~{\em 32},
  245--268.

\bibitem[\protect\citeauthoryear{Bedford and Cooke}{Bedford and
  Cooke}{2002}]{BedfordCooke2002}
Bedford, T. and R.~Cooke (2002).
\newblock Vines - a new graphical model for dependent random variables.
\newblock {\em Annals of Statistics\/}~{\em 30(4)}, 1031--1068.

\bibitem[\protect\citeauthoryear{Bollen}{Bollen}{1989}]{Bollen1989}
Bollen, K.~A. (1989).
\newblock {\em Structural Equations with Latent Variables\/} (1st ed.).
\newblock John Wiley and Sons, Chicester.

\bibitem[\protect\citeauthoryear{Brechmann, Czado, and Aas}{Brechmann
  et~al.}{2012}]{brechmann-etal}
Brechmann, E., C.~Czado, and K.~Aas (2012).
\newblock Truncated regular vines in high dimensions with application to
  financial data.
\newblock {\em Canadian Journal of Statistics\/}~{\em 40}, 68--85.

\bibitem[\protect\citeauthoryear{Brechmann and Joe}{Brechmann and
  Joe}{2014}]{BrechmannJoe2014}
Brechmann, E.~C. and H.~Joe (2014).
\newblock Parsimonious parameterization of correlation matrices using truncated
  vines and factor analysis.
\newblock {\em Computational Statistics \& Data Analysis\/}~{\em 77}, 233--251.

\bibitem[\protect\citeauthoryear{Brechmann and Schepsmeier}{Brechmann and
  Schepsmeier}{2013}]{cdvine2013}
Brechmann, E.~C. and U.~Schepsmeier (2013).
\newblock Modeling {D}ependence with {C}- and {D}-{V}ine {C}opulas: The {R}
  package {CDVine}.
\newblock {\em Journal of Statistical Software\/}~{\em 52\/}(3), 1--27.

\bibitem[\protect\citeauthoryear{Di{\ss}mann, Brechmann, Czado, and
  Kurowicka}{Di{\ss}mann et~al.}{2013}]{dissmann-etal}
Di{\ss}mann, J., E.~Brechmann, C.~Czado, and D.~Kurowicka (2013).
\newblock Selecting and estimating regular vine copulae and application to
  financial returns.
\newblock {\em Computational Statistics and Data Analysis\/}~{\em 52\/}(1),
  52--59.

\bibitem[\protect\citeauthoryear{Friedman, Hastie, and Tibshirani}{Friedman
  et~al.}{2008}]{glasso}
Friedman, J., T.~Hastie, and R.~Tibshirani (2008).
\newblock Sparse inverse covariance estimation with the graphical lasso.
\newblock {\em Biostatistics\/}~{\em 9\/}(3), 432.

\bibitem[\protect\citeauthoryear{Friedman, Hastie, and Tibshirani}{Friedman
  et~al.}{2010}]{glmnet}
Friedman, J., T.~Hastie, and R.~Tibshirani (2010).
\newblock Regularization {P}aths for {G}eneralized {L}inear {M}odels via
  {C}oordinate {D}escent.
\newblock {\em Journal of Statistical Software\/}~{\em 33\/}(1), 1--22.

\bibitem[\protect\citeauthoryear{Frommlet, Chakrabarti, Murawska, and
  Bogdan}{Frommlet et~al.}{2011}]{Frommletetal2011}
Frommlet, F., A.~Chakrabarti, M.~Murawska, and M.~Bogdan (2011).
\newblock Asymptotic {B}ayes optimality under sparsity for generally
  distributed effect sizes under the alternative.
\newblock Technical report.

\bibitem[\protect\citeauthoryear{Gruber and Czado}{Gruber and
  Czado}{2015a}]{GruberCzado20152}
Gruber, L. and C.~Czado (2015a).
\newblock Bayesian model selection of regular vine copulas.
\newblock {\em Preprint\/}.

\bibitem[\protect\citeauthoryear{Gruber and Czado}{Gruber and
  Czado}{2015b}]{GruberCzado2015}
Gruber, L. and C.~Czado (2015b).
\newblock Sequential bayesian model selection of regular vine copulas.
\newblock {\em Bayesian Analysis\/}~{\em 10}, 937--963.

\bibitem[\protect\citeauthoryear{Hastie, Tibshirani, and Wainwright}{Hastie
  et~al.}{2015}]{HastieTibshiraniWainwright2015}
Hastie, T., R.~Tibshirani, and M.~Wainwright (2015).
\newblock {\em Statistical Learning with Sparsity The Lasso and
  Generalizations}.
\newblock Boca Raton: CRC Press.

\bibitem[\protect\citeauthoryear{Hoyle}{Hoyle}{1995}]{Hoyle1995}
Hoyle, R.~H. (1995).
\newblock {\em Structural Equation Modeling\/} (1st ed.).
\newblock SAGE Publications, Thousand Oaks.

\bibitem[\protect\citeauthoryear{Kaplan}{Kaplan}{2009}]{Kaplan2009}
Kaplan, D. (2009).
\newblock {\em Structural Equation Modeling: Foundations and Extensions\/} (2nd
  ed.).
\newblock SAGE Publications, Thousand Oaks.

\bibitem[\protect\citeauthoryear{Kurowicka and Cooke}{Kurowicka and
  Cooke}{2006}]{KurowickaCooke2006}
Kurowicka, D. and R.~Cooke (2006).
\newblock {\em Uncertainty Analysis and High Dimensional Dependence
  Modelling\/} (1st ed.).
\newblock John Wiley \& Sons, Ltd, Chicester.

\bibitem[\protect\citeauthoryear{Kurowicka and Joe}{Kurowicka and
  Joe}{2011}]{kuro:joe:2010}
Kurowicka, D. and H.~Joe (2011).
\newblock {\em Dependence Modeling - Handbook on Vine Copulae}.
\newblock Singapore: World Scientific Publishing Co.

\bibitem[\protect\citeauthoryear{Meinshausen and B{\"u}hlmann}{Meinshausen and
  B{\"u}hlmann}{2006}]{meinshausen2006}
Meinshausen, N. and P.~B{\"u}hlmann (2006, 06).
\newblock High-dimensional graphs and variable selection with the {L}asso.
\newblock {\em Ann. Statist.\/}~{\em 34\/}(3), 1436--1462.

\bibitem[\protect\citeauthoryear{M{\"u}ller and Czado}{M{\"u}ller and
  Czado}{2016}]{MuellerCzado2016}
M{\"u}ller, D. and C.~Czado (2016).
\newblock Representing {S}parse {G}aussian {DAGs} as {S}parse {R-vines}
  {A}llowing for {N}on-{G}aussian {D}ependence.
\newblock {\em arXiv preprint arXiv:1604.04202\/}.

\bibitem[\protect\citeauthoryear{Schepsmeier, St{\"o}ber, Brechmann, Graeler,
  Nagler, and Erhardt}{Schepsmeier et~al.}{2016}]{VineCopula}
Schepsmeier, U., J.~St{\"o}ber, E.~C. Brechmann, B.~Graeler, T.~Nagler, and
  T.~Erhardt (2016).
\newblock {\em VineCopula: Statistical Inference of Vine Copulas}.
\newblock R package version 2.0.6.

\bibitem[\protect\citeauthoryear{Schwarz}{Schwarz}{1978}]{Schwarz1978}
Schwarz, G. (1978).
\newblock Estimating the dimension of a model.
\newblock {\em The Annals of Statistics\/}~{\em 6\/}(2), 461--464.

\bibitem[\protect\citeauthoryear{Sklar}{Sklar}{1959}]{Sklar1959}
Sklar, A. (1959).
\newblock Fonctions d\'e repartition \'a n dimensions et leurs marges.
\newblock {\em Publ. Inst. Stat. Univ. Paris\/}~{\em 8}, 229--231.

\bibitem[\protect\citeauthoryear{St\"ober, Joe, and Czado}{St\"ober
  et~al.}{2013}]{stoeber-vines}
St\"ober, J., H.~Joe, and C.~Czado (2013).
\newblock Simplified pair copula constructions-limitations and extensions.
\newblock {\em Journal of Multivariate Analysis\/}~{\em 119\/}(0), 101 -- 118.

\bibitem[\protect\citeauthoryear{Tibshirani}{Tibshirani}{1994}]{Tibshirani94regressionshrinkage}
Tibshirani, R. (1994).
\newblock Regression {S}hrinkage and {S}election {V}ia the {L}asso.
\newblock {\em Journal of the Royal Statistical Society, Series B\/}~{\em 58},
  267--288.

\bibitem[\protect\citeauthoryear{Zou and Hastie}{Zou and
  Hastie}{2005}]{ZouHastie2005}
Zou, H. and T.~Hastie (2005).
\newblock Regularization and variable selection via the elastic net.
\newblock {\em Journal of the Royal Statistical Society: Series B (Statistical
  Methodology)\/}~{\em 67\/}(2), 301--320.

\end{thebibliography}
\addcontentsline{toc}{section}{REFERENCES}

\newpage

\appendix

\setcounter{table}{0}
\renewcommand{\thetable}{A\arabic{table}}
\renewcommand{\thefigure}{A\arabic{figure}}
\setcounter{page}{1}

\begin{appendix}
	\section{Cross validation for the Lasso}\label{sec:app:cv}
	Assume a setup as introduced in the section \ref{sec:lasso}. We divide the total data set of $n$ observations into $k > 1$ randomly chosen subsets $K_1,\dots,K_k$ such that $\bigcup_{i=1}^k~K_i = n$. We obtain $k$ training data sets $S_{tr} = n \setminus K_m$ and corresponding test data sets $S_{te} = K_m$, $m=1,\dots,k$. Then, the coefficient vector $\widehat{\bm{\varphi}}_\ell = \left(\widehat{\varphi}_1^\ell,\dots,\widehat{\varphi}_p^\ell\right) \in \mathbb{R}^p$ is estimated for various $\lambda_\ell,\ \ell=1,\dots,L$ on each of the $k$ training sets. Now we use these $L$ coefficient vectors to predict for each test data set the values 
	\begin{equation*}
		{\widehat{y}}_i^\ell = \sum_{j=1}^p~\widehat{\varphi}_{j}^\ell x_{i,j},\ i \in K_m,\ m = 1,\dots,k,\ \ell = 1,\dots,L.
	\end{equation*}
	For these values, we also know the true values $y_i$, $i \in K_m$, $m=1,\dots,k$. Thus, we can calculate the \textit{mean squared prediction error} for this pair of training and test data:
	\begin{equation*}
		\delta_m^\ell = \frac{1}{|K_m|} \sum_{i \in K_m}~\left(y_i - \widehat{y}_i^\ell\right)^2,\ m = 1,\dots,k.
	\end{equation*}
	Since we have $k$ pairs of training and test data, we obtain an estimate for the prediction error for each of the $L$ values of $\lambda_\ell,\ \ell=1,\dots,L$ by averaging:
	\begin{equation*}
		\Delta_\ell = \frac{1}{k} \sum_{m = 1}^k~\delta_m^\ell,\ \ell = 1,\dots,L.
	\end{equation*}
	Next, consider the dependence between $\lambda_\ell,\ \ell=1,\dots,L$ and the corresponding error $\Delta_\ell$. A natural choice is to select $\lambda = \lambda_\ell$ such that $\Delta_\ell$ is minimal in $\left(\Delta_1,\dots,\Delta_L\right)$, we denote this by $\lambda_{min}^{CV}$. Alternatively, we choose $\lambda_\ell$ such that it is at least in within one-standard error of the minimum, denote $\lambda_{1se}^{CV}$. For both types of cross validation methods, see \citet{glmnet} or \citet[p.\ 13]{HastieTibshiraniWainwright2015}.
	
	\section{Additional results of the simulation study}\label{sec:app:simstudy}
	\begin{figure*}[ht]
		\centering
		\includegraphics[width=0.32\textwidth, trim={0.1cm 0.1cm 0.1cm 0.1cm},clip]{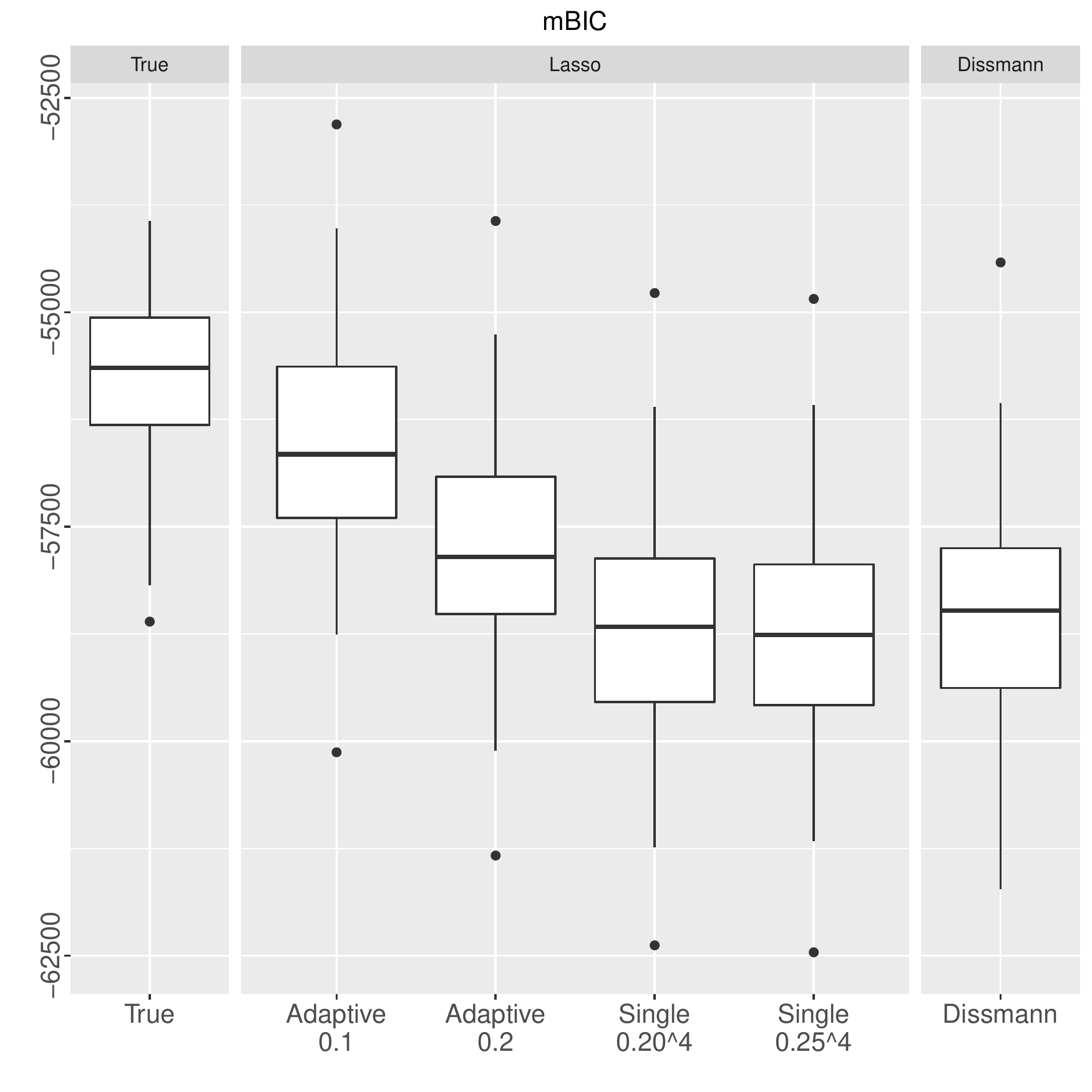}
		\includegraphics[width=0.32\textwidth, trim={0.1cm 0.1cm 0.1cm 0.1cm},clip]{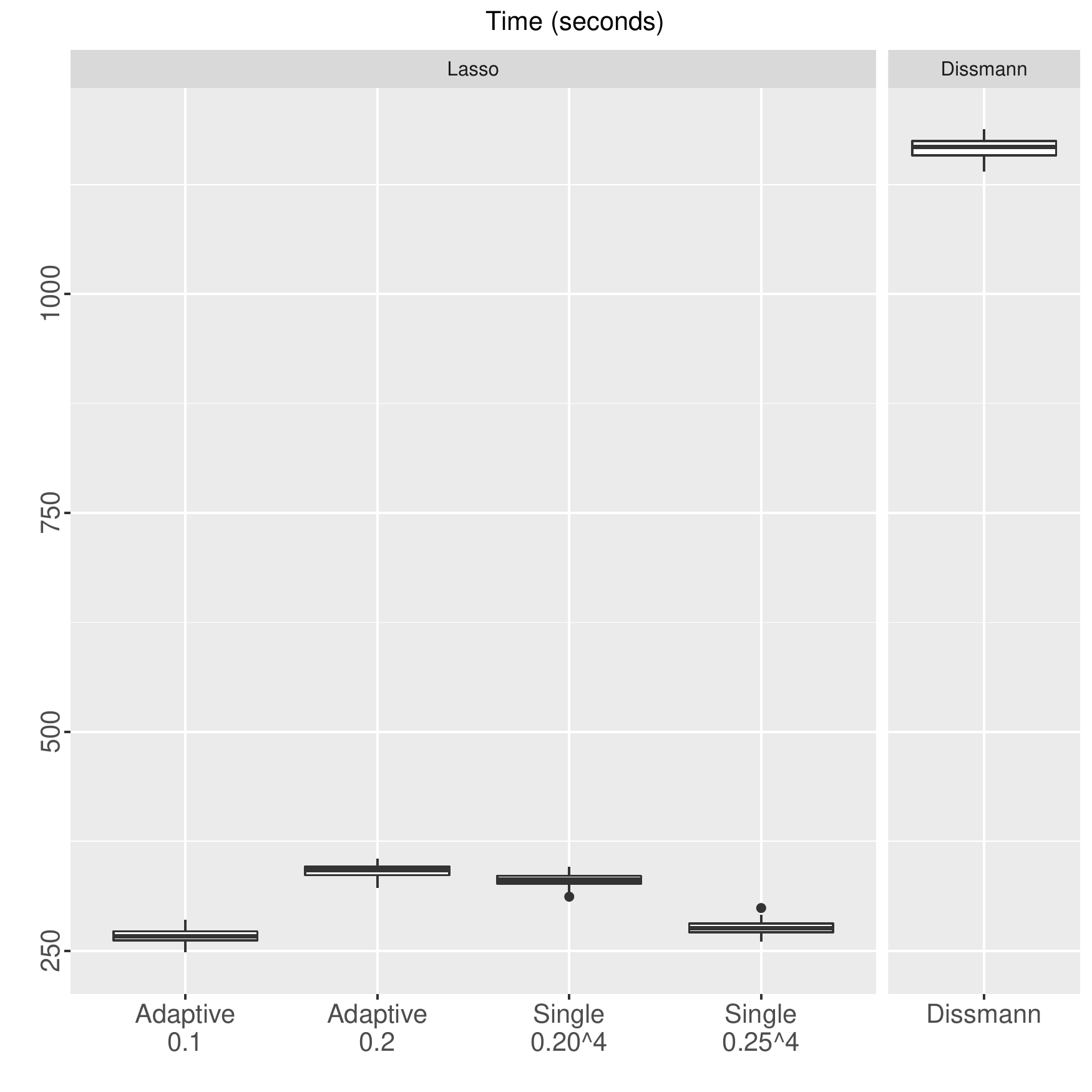}
		\includegraphics[width=0.32\textwidth, trim={0.1cm 0.1cm 0.1cm 0.1cm},clip]{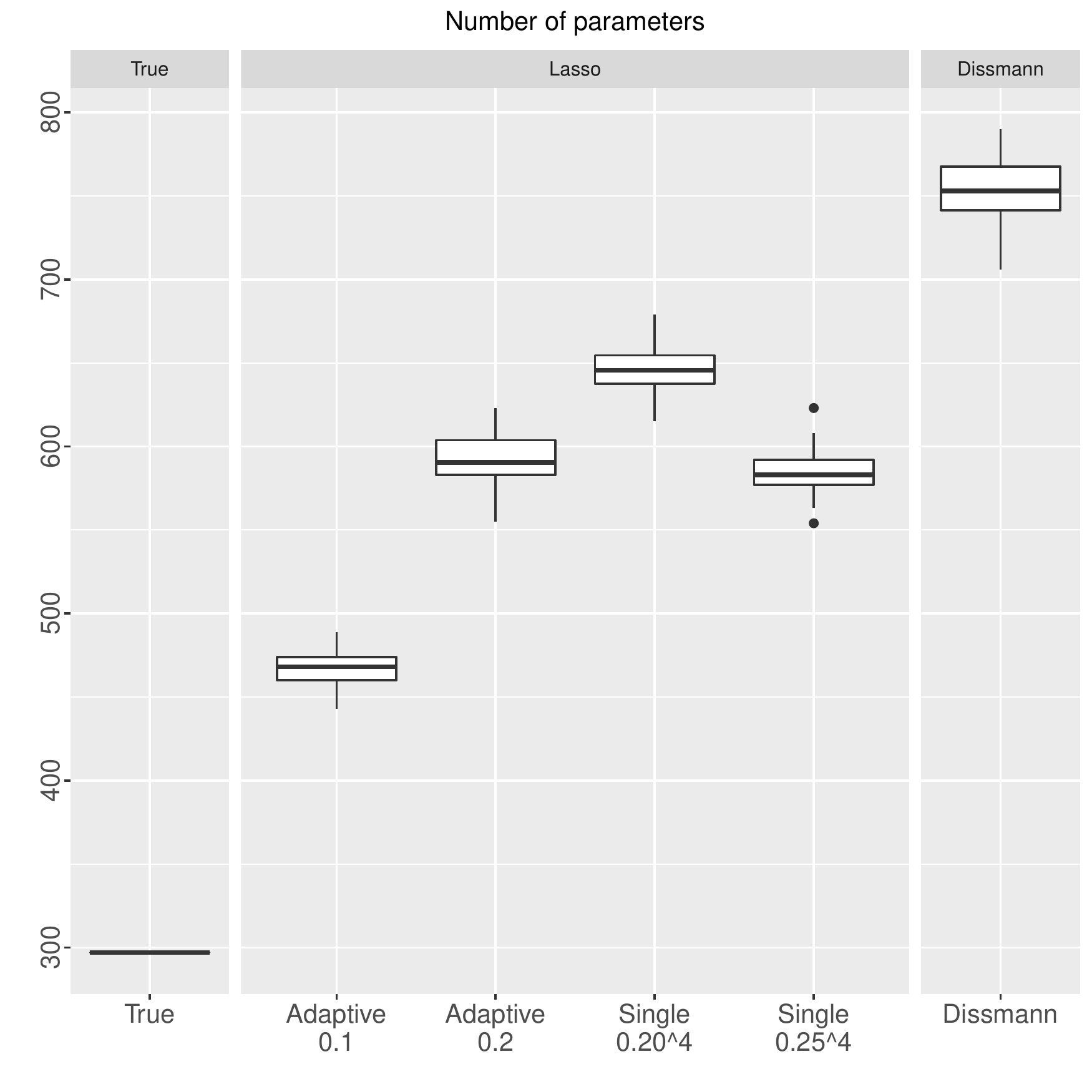}
		\caption{Scenario $S_2$: Comparison of Lasso approach for fixed and flexible thresholding with true model and Dissmann's algorithm considering mBIC, computation time in seconds and number of parameters on $50$ replications (from left to right).}
		\label{fig:simstudy:results2}
	\end{figure*}
	
	\begin{figure*}[ht]
		\centering
		\includegraphics[width=0.32\textwidth, trim={0.1cm 0.1cm 0.1cm 0.1cm},clip]{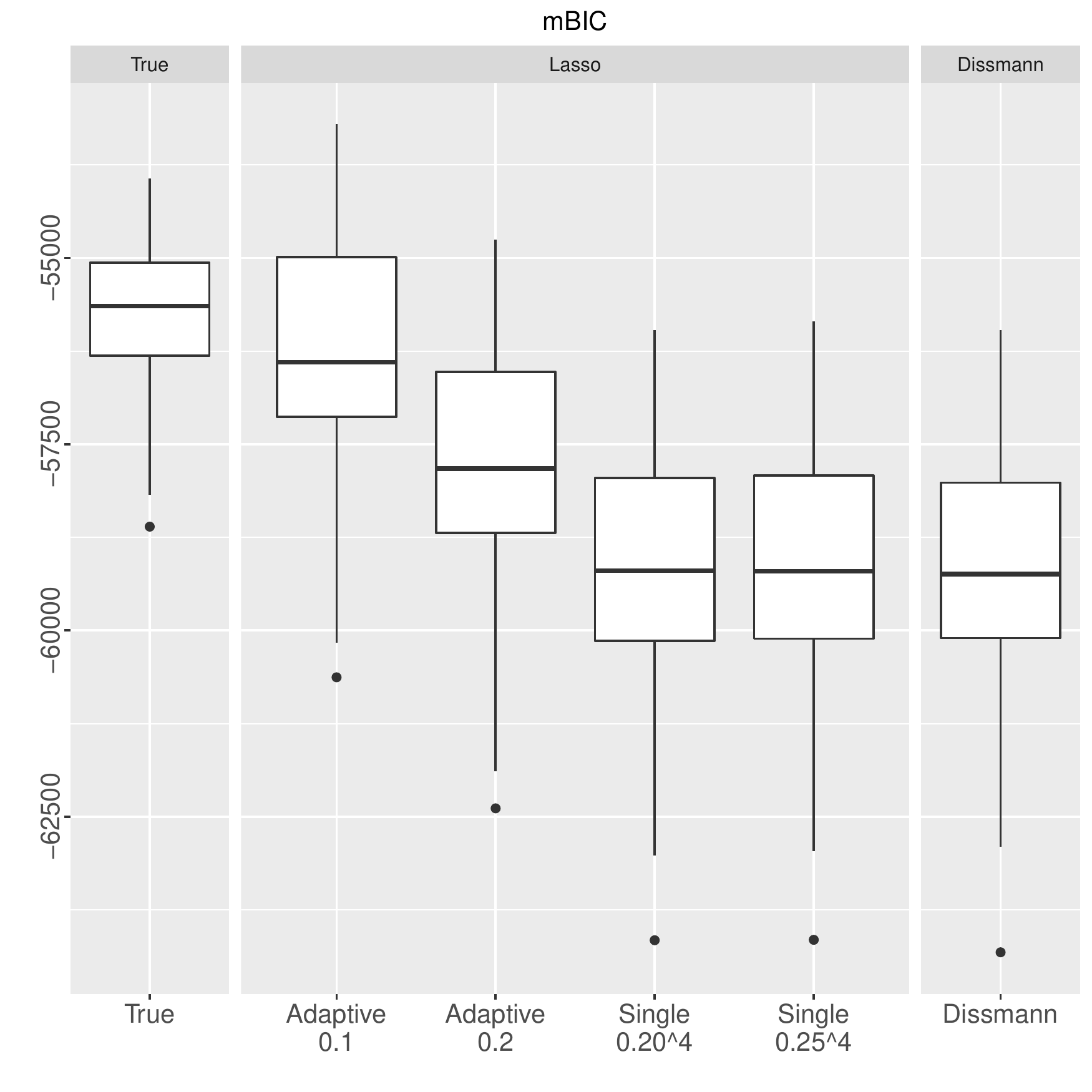}
		\includegraphics[width=0.32\textwidth, trim={0.1cm 0.1cm 0.1cm 0.1cm},clip]{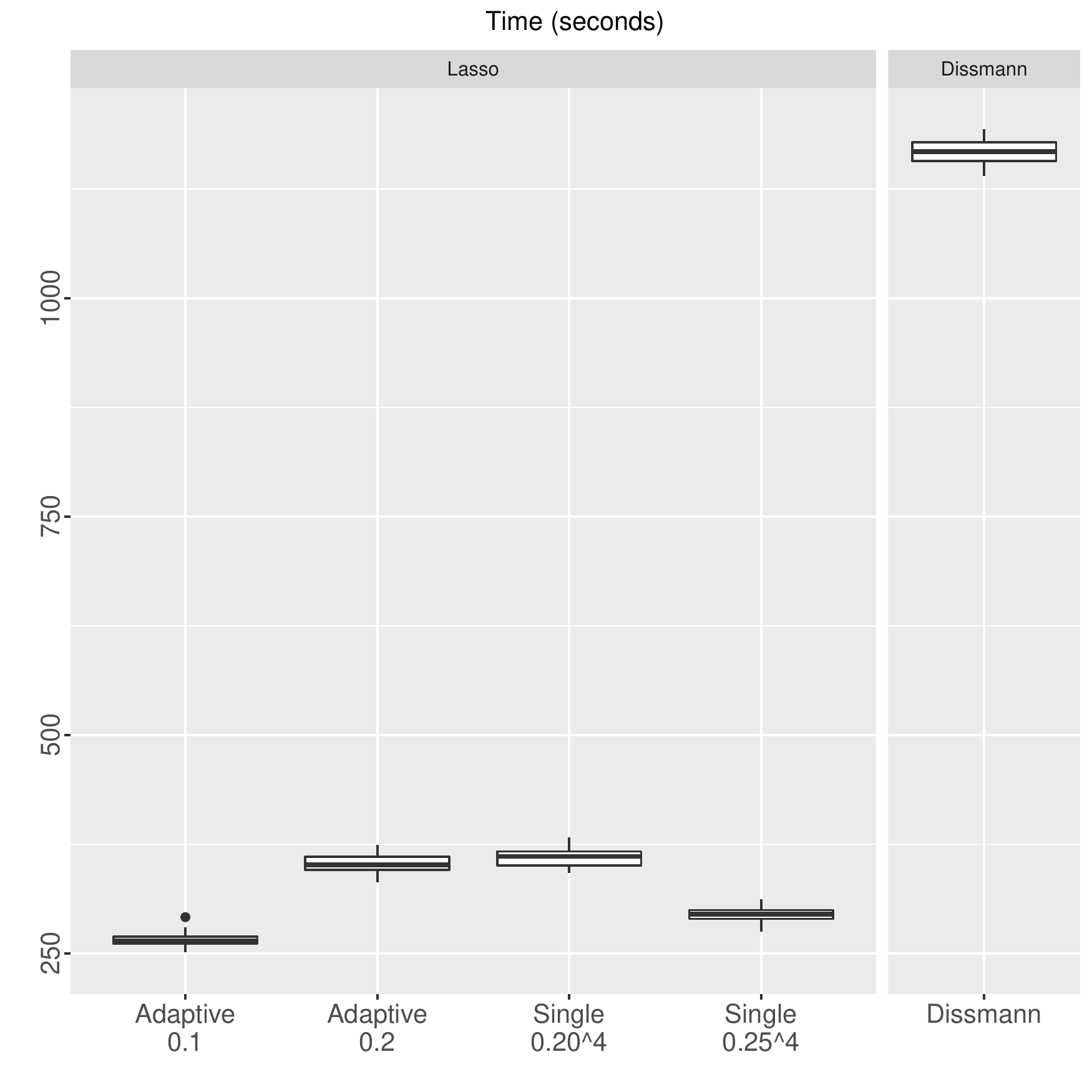}
		\includegraphics[width=0.32\textwidth, trim={0.1cm 0.1cm 0.1cm 0.1cm},clip]{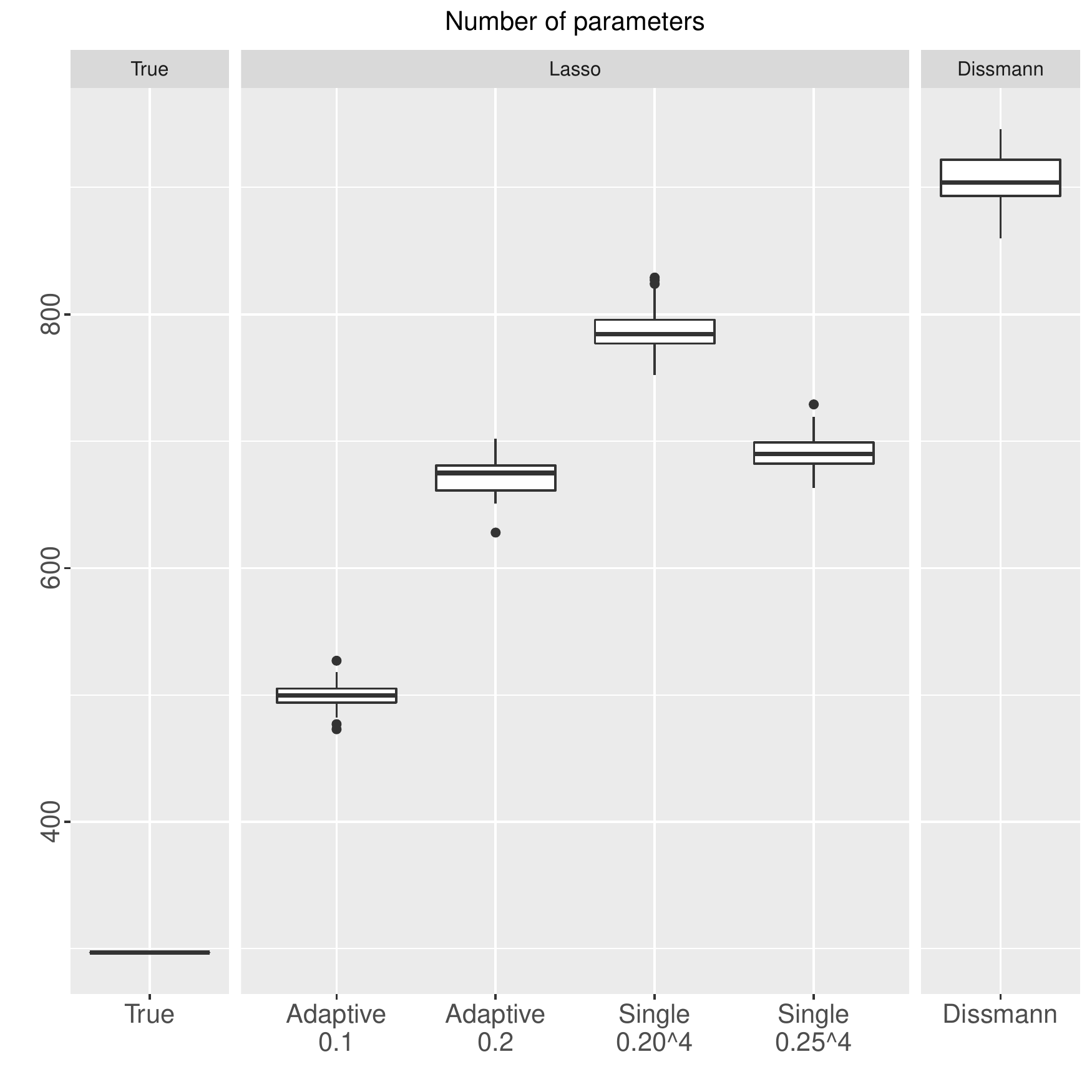}
		\caption{Scenario $S_3$: Comparison of Lasso approach for fixed and flexible thresholding with true model and Dissmann's algorithm considering mBIC, computation time in seconds and number of parameters on $50$ replications (from left to right).}
	\end{figure*}
	
\end{appendix}

\end{document}